\documentclass[journal=jpcafh,manuscript=article]{achemso}
\usepackage{longtable}
\usepackage{multirow}
\usepackage{amsmath}
\usepackage{subfigure}
\usepackage[usenames,dvipsnames]{xcolor}
\usepackage{soul}
\usepackage{bm}

\usepackage{xr}
\usepackage{amssymb}
\usepackage{siunitx}
\usepackage{threeparttable}
\usepackage{multicol} \usepackage{wrapfig} \usepackage{enumitem}
\usepackage{bm} \usepackage{outlines} %for itemize with subitems
\usepackage{booktabs}
\usepackage{graphicx}
\usepackage[normalem]{ulem}
\usepackage{hyperref}

\usepackage{lineno}
\usepackage[version=4]{mhchem}
\usepackage[section]{placeins}
\author{Abhirami Vijayakumar}
\affiliation{Department of Chemistry, University of Basel,
  Klingelbergstrasse 80, CH-4056 Basel,
  Switzerland}\altaffiliation{These authors contributed equally}

\author{Raidel Martin Barrios}
\affiliation{Department of Chemistry, University of Basel,
  Klingelbergstrasse 80, CH-4056 Basel,
  Switzerland}\altaffiliation{These authors contributed equally}

\author{Meenu Upadhyay, JingChun Wang} \affiliation{Department of
  Chemistry, University of Basel, Klingelbergstrasse 80, CH-4056
  Basel, Switzerland}

\author{Andre Staudte} \affiliation{National Research Council Canada,
  100 Sussex Drive, Ottawa ON K1A0R6 Canada}

\author{Albert Stolow} \affiliation{Department of Chemistry and
  Biomolecular Sciences, Department of Physics, University of Ottawa
  and National Research Council Canada, Ottawa Canada}
\email{astolow@uottawa.ca}

\author{Markus Meuwly}\email{m.meuwly@unibas.ch}
\affiliation{Department of Chemistry, University of Basel,
  Klingelbergstrasse 80, CH-4056 Basel, Switzerland}

\title[]{Observing Phase Space Evolution During Unimolecular Decay}

\begin{document}
\date{\today}

\begin{abstract}
Direct time-resolved observation of large amplitude dynamics during
ground state statistical unimolecular decay remains an experimental
challenge. Here, an experimental Time-Resolved Coulomb Explosion
Imaging (TR-CEI) study (J. Chem. Phys. 151, 174301, 2019) of the
canonical unimolecular reaction NO$_2$ $\rightarrow$ NO($^2\Pi, v=0,
J)$ + O($^3P_j$) at threshold dissociation energies is modeled and
analyzed. Quasi-classical trajectory (QCT) simulations on an accurate
and validated ground state potential energy surface (PES), with energy
and angular momentum constraints from known product state
distributions, are used to characterize the dissociation
dynamics. TR-CEI projects suddenly, via ultrafast Strong Field
Ionization (SFI), the evolving neutral ground state wavepacket onto
the repulsive doubly charged cationic state which then impulsively
fragments into the NO$^{+}$ + O$^{+}$ ion pair. Within the Coulomb
recoil approximation, which applies at longer NO--O (Jacobi)
separations, the ion pair CEI-energy is inversely proportional to the
scalar distance between the neutral ground state NO and O moieties at
the moment of SFI, thus providing a direct measure of the evolving
average NO--O distance during unimolecular decay. Results from QCT
simulations agree favourably with the TR-CEI measurements. Phase space
evolution, recurrence time distributions and Lyapunov exponent
analysis are presented which support the conclusion that TR-CEI
directly observes phase space evolution during unimolecular decay. For
the present system, the time scale for the phase space distribution to
achieve statisticality is estimated to be 2 to 3 ps.
\end{abstract}

\section{Introduction}
The characterization of ground state unimolecular photodissociation
dynamics near dissociation thresholds remains a challenging problem
due to the extreme density of states, large amplitude motion,
overlapping resonances and long decay times. A commonly accepted
treatment is that of statistical unimolecular decay: the available
microcanonical phase space is rapidly filled on a time scale short
compared to that of the reaction, requiring that intramolecular
vibrational energy redistribution (IVR) be both fast and complete on
the reaction time scale.\cite{baer1996unimolecular} This "ergodicity"
assumption greatly simplifies the calculation of reaction rates, but
it remains a fundamental question, whether the dynamics are truly or
only apparently statistical.\cite{remacle:1991,remacle:1998} Apparent
agreement with statistical theories may depend on the chosen
observable, such as a state-resolved rate or product state
distributions. Since the filling of phase space takes time, the
dynamics cannot be "instantaneously" statistical. Different
observables may exhibit different timescales for the onset of
statisticality. It is therefore desirable to develop experimentally
accessible observables which are not only sensitive to the final
products, but directly probe their phase space evolution during
unimolecular decay.\\

\noindent
A standard approach to preparing a microcanonical ground state
distribution is via photoabsorption to an electronically excited state
which undergoes extremely rapid (i.e., $< 100$ fs) internal conversion
to the ground electronic state, thus converting photon energy into
ground state vibrational energy - a so-called ``hot'' molecule which
then proceeds to dissociate on the ground state potential surface
(PES). In a time-resolved experiment, an ultrashort pump pulse is used
and a delayed ultrashort probe pulse generates an observable (e.g.,
photofragment yield). Here, the temporal region between initial
photoinduced ``hot'' ground state formation and the final appearance
of free products is considered. Can an observable be constructed which
probes this highly complex large amplitude motion region as the system
evolves towards a loose transition state and the subsequent emission
of free products? \\

\noindent
Time-Resolved Coulomb Explosion Imaging (TR-CEI) has emerged as a
novel probe of unimolecular decay
dynamics\cite{endo2020capturing,stolow:2019} and was recently applied
to the threshold photodissociation dynamics of
NO$_2$.\cite{stolow:2019} In that experiment, an ultrashort pump pulse
at 400 nm excited NO$_2$ to its optically bright $^2B_2$ state, an
internal energy corresponding to the NO$_2$ $\rightarrow$ NO($^2\Pi,
v=0, J)$ + O($^3P_j$) dissociation threshold of $D_0 = 25,128.56$
cm$^{-1}$ (397.954 nm). \cite{jost1996photodissociation} Ultrafast
internal conversion subsequently leads to formation of the ``hot''
electronic ground
state.\cite{arasaki:2007,arasaki:2010,forbes:2017,von2018conical} By
50-60 fs, a complex ground state wavepacket has formed, with
population transfer to the ground state largely complete by $\sim$ 250
fs. These non-adiabatic processes are the initial dynamics in the
formation of the ``hot'' ground state NO$_2$ molecule, designated here
as [NO$_2$]$^\nmid$, which is understood to subsequently undergo
strong IVR and very large amplitude motions. In the following, only
slightly longer time scales wherein NO$_2$ can be treated as a ``hot''
ground state molecule are considered.\\

\noindent
In TR-CEI, the SFI probe pulse projects the evolving neutral ground
state wavepacket onto the repulsive doubly charged cationic state
which then fragments into NO$^{+}$ + O$^{+}$ ion pairs. Within the
Coulomb recoil approximation, which becomes accurate at longer NO--O
distances, the ion pair recoil energy is inversely proportional to the
scalar distance (Jacobi coordinates) between the neutral ground state
NO and O moieties at the moment of SFI, thus providing a measure of
the evolving average NO--O distance during unimolecular decay. TR-CEI
is a potentially unique measure of large amplitude ground state phase
space evolution during statistical unimolecular decay. In the
following, the TR-CEI experiment is analyzed through a combination of
quasi-classical trajectory (QCT) simulations and detailed phase space
analysis.\\

\noindent
NO$_2$ is a well-suited system to the study of statistical
vs. non-statistical photodissociation dynamics because the NO$_2$
$\rightarrow$ NO+O channel is accessible near threshold, where small
changes in excess energy strongly affect fragmentation. State-specific
near-threshold experiments showed that individually prepared states
can exhibit distinct unimolecular decay behavior rather than perfectly
averaged statistical fragmentation.\cite{Miyawaki1993NO2Threshold}
Product-state-resolved measurements of NO fragments provide
rotational, vibrational, and translational distributions which
directly test statistical phase-space
predictions.\cite{Hunter1993NO2ProductDistributions} Time-resolved
unimolecular-decomposition experiments further connected fragment
observables to the evolving dissociation dynamics near the NO$_2$
$\rightarrow$ NO+O transition
region.\cite{Ionov1993NO2TransitionState} Since NO$_2$ is triatomic,
full-dimensional potential-energy surfaces and detailed dynamics
studies are feasible, while the dynamics remain nontrivial.  Schinke
and co-workers established NO$_2$ as a paradigm for comparing
classical trajectory calculations with RRKM and statistical adiabatic
channel models on a global ground-state PES
\cite{Grebenshchikov1999NO2Statistical}.  This renders NO$_2$
especially useful for testing the statistical assumption that
intramolecular vibrational redistribution is indeed fast compared with
bond fission. NO$_2$ is an ideal case because the decades of
dedicated, ongoing experimental and theoretical work provide
information about state-resolved photofragmentation and time-resolved
imaging. This makes NO$_2$ microcanonical dynamics a stringent test of
whether the photofragmentation is statistical or if it retains mode-
or state-specific dynamical
memory.\cite{Grebenshchikov1999NO2Statistical,Hunter1993NO2ProductDistributions,Ding2019NO2COLTRIMS}\\

\noindent
A wide range of potential energy surfaces (PESs) for the reactive O+NO
system has been developed each tailored for different
applications. For example, the vibrational state-dependent cross
sections were calculated from dynamics
studies.\cite{Caridade2004,Duff1994,Ramachandran2000} One
PES\cite{Duff1994} for the $^2$A$'$ state was used a fit to electronic
structure calculations at the complete active space SCF (CASSCF)
level, followed by multireference contracted configuration interaction
and a modified Duijneveldt (11s6p) basis set.\cite{walch:1987} Another
PES was based on 1250 (for the $^2$A$'$) and 910 ($^4$A$'$) CASPT2
point calculations, with fitting to an analytical
function.\cite{Says2002} Such an approach was also used for the
$^2$A$''$ state.\cite{Gonzlez2001} Complementarily, a PES for the
$^2$A$'$ state using a diatomics in molecules (DIM) expansion with the
two-body terms was based on extended Hartree-Fock
calculations.\cite{varandas:2003} A 2-dimensional PES with the NO bond
length fixed at its equilibrium value of 2.176 a$_0$ was determined at
the icMRCI+Q and cc-pVQZ level of theory and represented as a cubic
spline.\cite{Ivanov2007} This work also presented a PES for the
$^2$A$''$ state. More recently, a double many body expansion fit to
1700 points at the MRCI/aug-cc-VQZ level of theory for the $^2$A$''$
state was carried out.\cite{Mota2012} In addition, QCT
simulations\cite{Caridade2004,CastroPalacio2014,Bose1997,Says2002}
were reported for the temperature dependent rate for the
N($^4$S)+O$_{2}$ $\rightarrow$ NO+O and its reverse reaction using
different PESs. Most recently, two machine learning-based PESs were
reported. One of them used CASPT2 reference data and was represented
as a many body permutation invariant polynomial\cite{varga:2021}
whereas the other used MRCI+Q calculations as the reference method,
together with a reproducing kernel Hilbert space (RKHS)
representation.\cite{MM.no2:2020,MM.rkhs:2017}\\

\noindent
The present study differs in emphasis from previous work which studied
the non-adiabatic dynamics on somewhat lower-level
CASSCF/cc-pVTZ-quality PESs between the X$^2$A$_1$/A$^2$B$_2$
states.\cite{arasaki:2007} There, an initial wavepacket on the ground
state was projected onto the electronically excited state and
propagated through the conical intersection (CI). After reaching this
crossing, one part of the wavepacket switched to the ground electronic
state whereas the remainder continues on the excited state PES. Upon
reaching the CI for a second time, again a fraction switches to the
ground state. It was reported that after $\sim 200$ fs most of the
excited state population transferred to the ground state, forming
[NO$_2$]$^\nmid$, from where dissociation into NO+O was followed in a
fashion comparable to the present work but no specific observable was
constructed which permits following the dynamics to long range. Hence,
differences between the present work and earlier simulations concern
(i) non-adiabatic effects, which were not included here since the
focus is on the hot ground state [NO$_2$]$^\nmid$ after all
non-adiabatic dynamics are complete, and (ii) the accuracy of the
underlying PES(s) which is higher in the present work.\\

\noindent
This manuscript is structured as follows. First, the theory and
simulation methods are presented, followed by characterization of the
low-energy collision process, final product rotational distributions
and the details of ensemble of reactive trajectories. The trajectory
data were used to model the TR-CEI experiments, permitting a direct
comparison of experiment with theory. A phase space analysis is
presented, including the sampling of individual trajectories,
recursion time distributions and Lyapunov exponents. Finally,
conclusions are drawn supporting the case that TR-CEI probes the
progressive loss of dynamical memory and the emergence of statistical
behavior in unimolecular dissociation of NO$_2$.\\

\section{Methods}

\subsection{The Potential Energy Surface}
Reference calculations for the $^2$A$'$ state of NO$_2$ connecting the
N($^4$S) + O$_2$(X$^3\Sigma^-_{\rm g})$ $\rightarrow$ O($^3$P) +
NO(X$^2\Pi)$ states were carried out at the MRCI+Q/aug-cc-pVTZ level
of theory, based on CASSCF calculations including all valence orbitals
in the active space. For the O+NO and N+O$_2$ channels, the grid
contained 7280 and 3920 geometries, respectively. The individual
channels (O$_{\rm A}$ + NO$_{\rm B}$, O$_{\rm B}$ + NO$_{\rm A}$, and
N + O$_{\rm A}$O$_{\rm B}$) were mixed using a distance-dependent
exponential switching\cite{MM.n2o:2020} for which a separate mixing
data set was constructed. For on- and off-grid points the $r^2 > 0.99$
with corresponding RMSE-values of 0.02 eV (0.5 kcal/mol) and 0.03 eV
(0.7 kcal/mol), respectively.\cite{MM.no2:2020} Details about other
PESs for the N+O$_2$ system are provided in the supporting
information.\\

\subsection{Quasi-Classical Trajectory Simulations}
The QCT method used in this work has been thoroughly described in the
literature.\cite{karplus:1965,truhlar:1979,MM.no2:2020,MM.co2:2021} In
this approach, Hamilton’s coupled differential equations of motion are
integrated using the fourth-order Runge--Kutta method in reactant
Jacobi coordinates with a time step of $\Delta t = 0.05$~fs, ensuring
conservation of total energy to within $10^{-6}\,E_\mathrm{h}$ per
trajectory. After adequate time propagation by monitoring the
internuclear distances, the momenta and positions are transformed to
the appropriate coordinate system (i.e., product Jacobi coordinates if
a product is formed). In order to simulate the TR-CEI pump-probe
experiment, a procedure was developed for generating the initial
($\tau=0$) energized NO$_2$ ground state configuration with internal
energy and angular momenta constraints. This required a
``synchronization'' of the trajectories, as described below.\\

\noindent
A total of 500\,000 quasi-classical trajectories were computed on the
NO$_2$ PES\cite{MM.no2:2020} for the initial rovibrational state
$(v=0,\, j=1)$. The initial reactant separation for the O + NO
collision was set to 22~$a_0$. The translational energy was equivalent
to a temperature of 50~K to generate the ``hot'' parent species
[NO$_2$]$^\nmid$. For this, the impact parameter $b$ was selected
using stratified sampling\cite{truhlar:1979,Bender:2015,Koner2018} in
the range $0 \le b \le b_{\rm max}$. To determine $b_{\rm max}$, the
opacity function for the exchange reaction $\mathrm{O_A +
  NO_B}~(v=0,\, j=1) \longrightarrow \mathrm{O_B + NO_A}$ was
determined, see Figure \ref{sifig:opacity}. Possible elementary
processes used for classification included elastic, inelastic and atom
exchange trajectories. In addition to atom exchange (see above), a
collision was classified as elastic/inelastic if the final NO-state
was $(v'=0,\, j'=1)$, and $(v',\, j') \neq (0,\,1)$, respectively.\\

\noindent
The final angular momentum $\mathbf{j}' = \mathbf{q}' \times
\mathbf{p}'$ of the product diatomic species was obtained from the
final position $\mathbf{q}'$ and momentum $\mathbf{p}'$. To determine
the rotational quantum number $j'$ as a real value, the following
quadratic expression was solved\cite{truhlar:1979,karplus:1965}:
\begin{equation}
j' = -\frac{1}{2} 
     + \frac{1}{2} 
       \left( 
         1 + 4\,\frac{\mathbf{j}' \cdot \mathbf{j}'}{\hbar^{2}} 
       \right)^{1/2}.
\label{eq:rotational_quantum}
\end{equation}

The vibrational quantum number ($\nu'$) of the final diatomic species
was subsequently determined as a non-integer value according
to\cite{truhlar:1979,karplus:1965}:
\begin{equation}
\nu' = -\frac{1}{2} 
       + \frac{1}{\pi \hbar}
         \int_{r^-}^{r^+}
         \left\{
            2\mu \left( E_{\mathrm{int}} - V(r) \right)
            - \frac{\mathbf{j} \cdot \mathbf{j}}{2 m r^{2}}
         \right\}^{1/2}
         \, dr .
\label{eq:vib_quantum}
\end{equation}
where $r$ denotes the diatomic bond length, $r^{+}$ and $r^{-}$
represent the turning points of the diatomic species on the effective
potential corresponding to rotational state $j'$ for internal energy
$E_{\mathrm{int}}$, $\mu$ is the reduced mass, and $V(r)$ is the
diatomic potential energy curve.\\

\noindent
{\it Simulations using measured $P(j)$:} In addition to standard
Boltzmann-type initial conditions for the NO-angular momentum $j$,
simulations were also performed by using experimentally measured
$P(j)$ from photofragment yield spectroscopy.\cite{reisler:1994} The
impact parameter $b$ for each trajectory was computed according to $ b
= \hbar \sqrt{\frac{j(j+1)}{2 \mu E}}$ where $j$ is the rotational
quantum number sampled from the probability distribution reported in
Reisler's work, $\mu$ is the reduced mass defined as the mass of the
oxygen atom divided by the total mass of the nitrogen and oxygen
atoms, and $E$ is the translational energy. The translational energy
$E_t$ was sampled from a Gaussian distribution centered at $E_t =
0.00431$ eV with a width of $\sigma(E_t) = 0.0008$ eV, representing
the energy spread. The number of initial conditions assigned to each
$j$ value was proportional to the experimentally measured probability
in the $P(j)$ distribution.\\

\subsection{Analysis}
In order to simulate the TR-CEI pump-probe experiments, a ``start
time'' was required for the creation of the energized ground state
[NO$_2$]$^\nmid$ molecule. Thus, for subsequent analyses, all
trajectories were first synchronized to have a common "time zero"
($\tau = 0$), which was defined as the instance at which $R = R^*$ for
the first time. Here, $R$ is the distance (Jacobi coordinates) between
the incoming oxygen atom O$_{\rm B}$ and the center of mass of the
diatomic NO$_{\rm A}$, and $R^* = 3.0$ a$_0$. The "lifetime"
$\tau_{\rm life}$ of metastable [NO$_2$]$^\nmid$ was then determined
according to Figure \ref{fig:sketch}.\\

\section{Results}
In the following, first the final state distribution after preparation
of the near-dissociative initial conditions and the dynamics in the
hot [NO$_2$]$^\nmid$ complex are analyzed. This is followed by the
calculation of the time- and energy-resolved NO$^+$ + O$^+$ Coulombic
recoil energy maps which correspond to the TR-CEI
experiments. Finally, the phase space trajectories are analyzed.\\

\subsection{Breakup and Final State Analysis}
The initial collision system was prepared in a metastable-dissociative
state which emulates the preparation of the ``hot'' ground state
molecule in the TR-CEI experiments. The opacity function shown in
Figure \ref{sifig:opacity} shows that the initial conditions chosen
cover the relevant collision parameters. In other words, the reaction
probability decays to zero for $b > 20$ a$_0$.\\

\begin{figure}[h!]
\includegraphics[width=\textwidth]{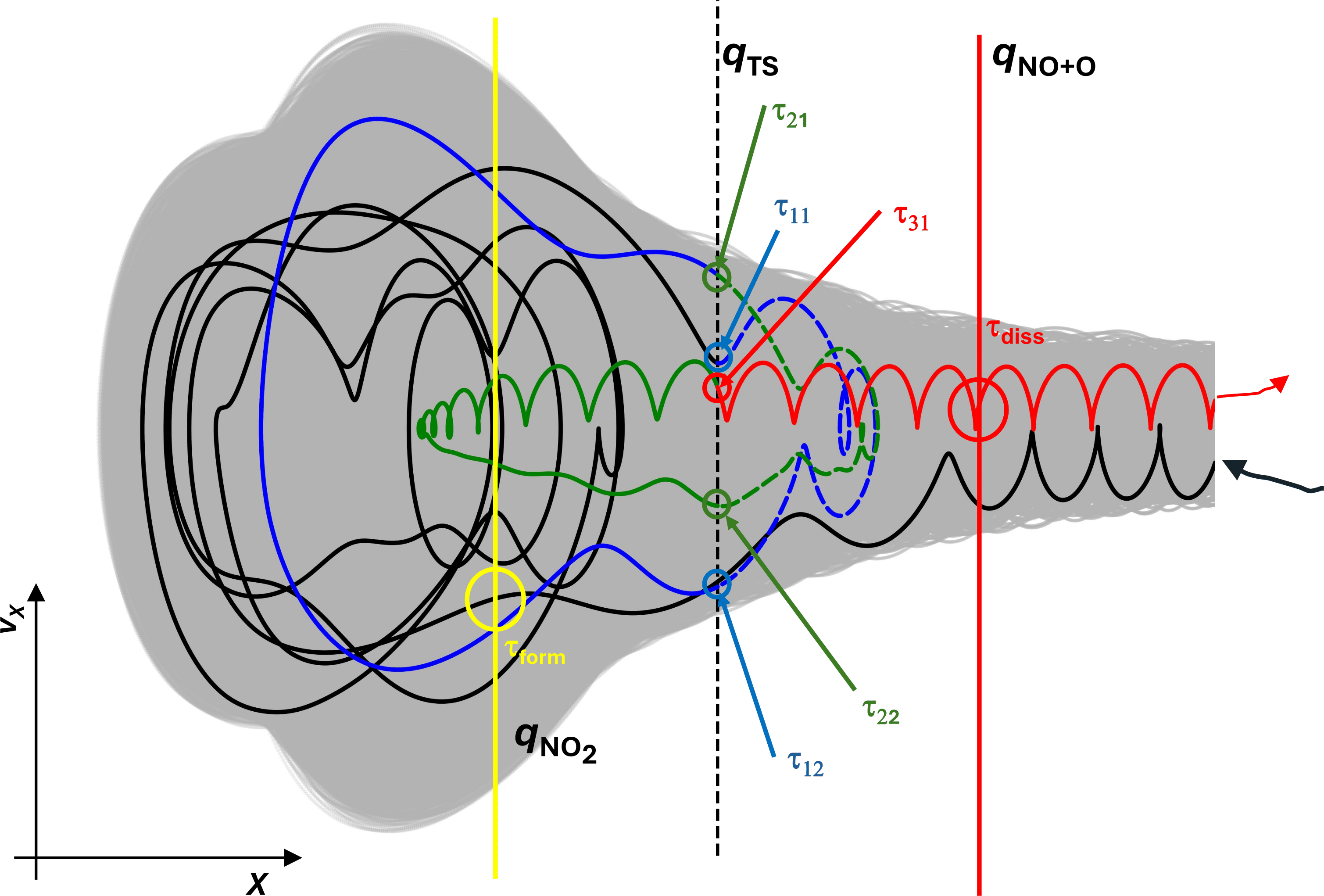}
\caption{Sketch for defining lifetimes and residence/recurrence
  times. The $x-$ and $y-$axes are an atom-atom separation (e.g. the
  NO$_{\rm A}$--O$_{\rm B}$ distance) and the corresponding
  velocities, respectively. Entering along the black line, reaching
  the value of $q_{\rm NO_2}$ (yellow line, first occurrence of $R <
  3$ a$_0$) indicates formation of [NO$_2$]$^\nmid$ at $\tau_{\rm
    form}$. This defines $\tau=0$ used for synchronization. Following
  the black line brings the system to $q_{\rm TS}$ (black dashed line,
  typically $R \sim 5$ a$_0$) at time $\tau_{11}$ (blue circle) for
  the first time point of the first crossing. Following the blue
  dashed line leads back to $q_{\rm TS}$ at time $\tau_{12}$ (blue
  circle) which is the second time point of the first crossing. The
  time difference $\Delta_1^{\rm rec} = \tau_{12} - \tau_{11}$ is the
  first recurrence time. Following the solid blue line brings the
  system again to $q = q_{\rm TS}$ at time $\tau_{21}$ (green
  circle). The time difference $\Delta_1^{\rm res} = \tau_{21} -
  \tau_{12}$ is the first residence time. Continuing along the green
  dashed line leads back to $q_{\rm TS}$ at time $\tau_{22}$ (green
  circle)) and $\Delta_2^{\rm rec} = \tau_{22} - \tau_{21}$ is the
  second recurrence time. After reaching $q_{\rm TS}$ again at
  $\tau_{31}$, this trajectory moves on to $q_{\rm NO+O}$ (red
  vertical line), and dissociates (red vertical line,
  $r_{12}+r_{23}+r_{13} > 25$ a$_0$) at $\tau_{\rm diss}$ by which the
  system has reached the asymptotic region. The sequence
  ${\Delta_1^{\rm rec},\Delta_2^{\rm rec},\cdots,\Delta_m^{\rm rec}}$
  constitutes the set of recurrence times and ${\Delta_1^{\rm
      res},\Delta_2^{\rm res},\cdots,\Delta_{m-1}^{\rm res}}$ are the
  residence times. The NO$_2$ lifetime is $\tau_{\rm life} = \tau_{\rm
    diss} - \tau_{\rm form}$.  It is clear that the values of the
  three distances $q_{\rm NO_2} < q_{\rm TS} < q_{\rm NO+O}$ are to
  some degree arbitrary. However, within reasonable limits, the final
  results do not change appreciably by varying them.}
\label{fig:sketch}
\end{figure}

\begin{figure}[h!]
    \centering \includegraphics[width=0.9\linewidth]{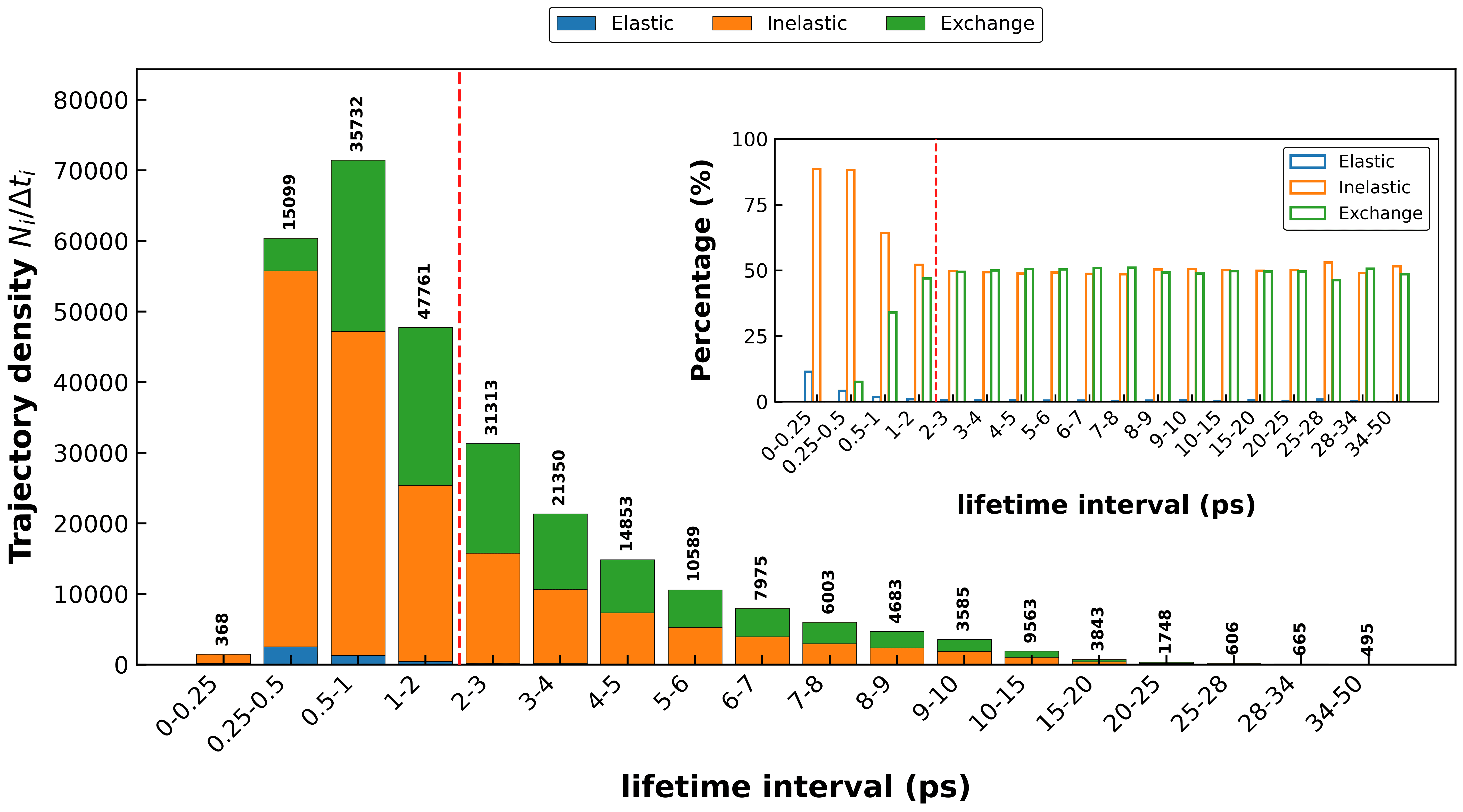}
    \caption{Distribution of lifetimes grouped into different time
      intervals. Here, the lifetime is defined as the difference
      between the dissociation time (when $r_{12}+r_{23}+r_{13} > 25$
      a$_0$) and the synchronization time (when $R = 3.0$~a$_0$ for
      the first time). The main panel shows the trajectory density,
      $N_i/\Delta t_i$, normalized by the width of each interval,
      $\Delta t_i$, since the intervals are non-equidistant, with
      contributions from elastic (blue), inelastic (orange), and
      exchange (green) processes. The numbers above the bars indicate
      the total number of trajectories in each $\tau_{\rm
        life}-$interval. The inset displays the relative percentage of
      each process within each lifetime interval. The red dashed line
      marks the boundary between short-lived ($\tau_{\rm life} < 2$
      ps) and longer-lived collision complexes.}
    \label{fig:poftlife}
\end{figure}

\noindent
The first property considered was the lifetime distribution $P(\tau)$
of the [NO$_2$]$^\nmid$ collision complex, shown in Figure
\ref{fig:poftlife}. As the final state is the O+NO channel, two
possibilities exist: elastic/inelastic scattering for which O$_{\rm
  A}$ + NO$_{\rm B}$ $\rightarrow$ O$_{\rm A}$ + NO$_{\rm B}$ or atom
exchange, characterized by O$_{\rm A}$ + NO$_{\rm B}$ $\rightarrow$
O$_{\rm B}$ + NO$_{\rm A}$. Analysis of the 500\,000 trajectories
showed that 216\,249 trajectories formed the hot [NO$_2$]$^\nmid$
complex after synchronization. Amongst these, 44\% resulted in
exchange, 55\% in inelastic scattering, and 1\% in elastic
scattering.\\

\begin{figure}[h!]
    \centering \includegraphics[width=0.88\linewidth]{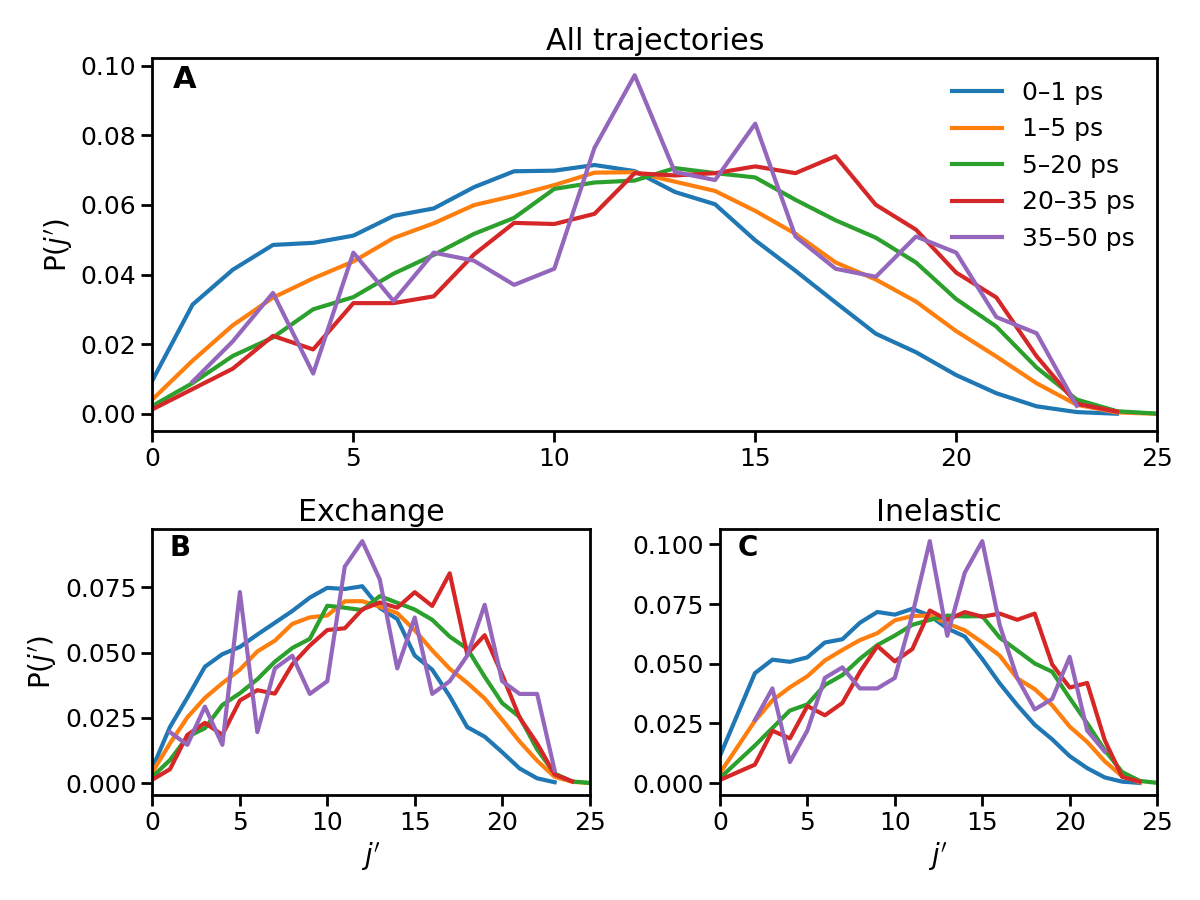}
    \caption{Final-state rotational distributions, $P(j')$, for
      trajectories with lifetimes in the range 0--50~ps. The impact
      parameter, $b$, was sampled using a stratified scheme over the
      interval $0 \leq b \leq 20$~a$_0$. Table \ref{tab:traj_counts}
      reports the number of trajectories for each lifetime interval
      and process considered: All (panel A), O-exchange (panel B), and
      inelastic (panel C). Figure \ref{sifig:pofj-fin} compares
      $P(j')$ for inelastic and exchange trajectories for each time
      interval.}
    \label{fig:pofj}
\end{figure}

\noindent
Next, the product rotational state distributions $P(j')$ were
analyzed. Because the collision energy is low (50 K) and the initial
NO$_{\rm B}$ is in $(v=0)$, vibrational excitation of the product
cannot occur: the vibrational quantum number is conserved by the
collision. Hence, the product state is fully characterized by the
relative kinetic energy of the O and NO fragments and $P(j')$. Figure
\ref{fig:pofj} presents the dependence of $P(j')$ on the
    [NO$_2$]$^\nmid$ lifetimes, ranging from $0 \leq \tau < 1$ ps
    (blue) to $35 \leq \tau < 50$ ps (indigo). Typically, with
    increasing lifetime, the maximum of $P(j')$ shifts to higher
    $j'-$values. This is most clearly seen when comparing $P(j')$ for
    $0 \leq \tau < 1$ ps (blue) with $20 \leq \tau < 35$ ps (red) for
    which $j'_{\rm max} = 10$ and $j'_{\rm max} = 16$,
    respectively. For the longest lifetimes, $P(j')$ is probably not
    yet converged: see Table \ref{tab:traj_counts} for the number of
    trajectories with specific lifetimes.\\

\noindent
Trajectories for which O-atom exchange occurs versus those which
involve an inelastic scattering process can be analyzed
separately. This is relevant because within the framework of QCT
simulations one expects that full statisticality would yield exchange
and inelastic scattering processes of equal probability. The
corresponding $P(j')$ are shown in Figures \ref{fig:pofj}B/C. Broadly
speaking, the distributions for each process agree qualitatively with
one another. However, there are some subtle differences. For example,
for inelastic collisions (blue), there is a first local maximum at $j'
= 4$ which is not present for atom-exchange trajectories. As the
number of trajectories for the two channels is sufficiently large
(39120 vs. 17984), it is unlikely that such differences are due to
statistical fluctuations. This is also confirmed by bootstrapping
which shows that the feature at $j' = 4$ for the inelastic channel is
"real", see also Figure \ref{sifig:pofj-fin}A. \\

\begin{table}[h!]
\centering
\caption{Number of trajectories contributing to the final-state
  rotational distributions in different lifetime intervals.}
\label{tab:traj_counts}
\begin{tabular}{lccccc}
\hline
 & 0--1 ps & 1--5 ps & 5--20 ps & 20--35 ps & 35--50 ps \\
\hline
All       & 51197 & 115277 & 46243 & 3082 & 432 \\
Exchange  & 13288 & 56087  & 23150 & 1518 & 205 \\
Inelastic & 36587 & 58278  & 22887 & 1550 & 227 \\
\hline
\end{tabular}
\end{table}
%\FloatBarrier

\begin{figure}[h!]
    \centering
    \includegraphics[width=1.0\linewidth]{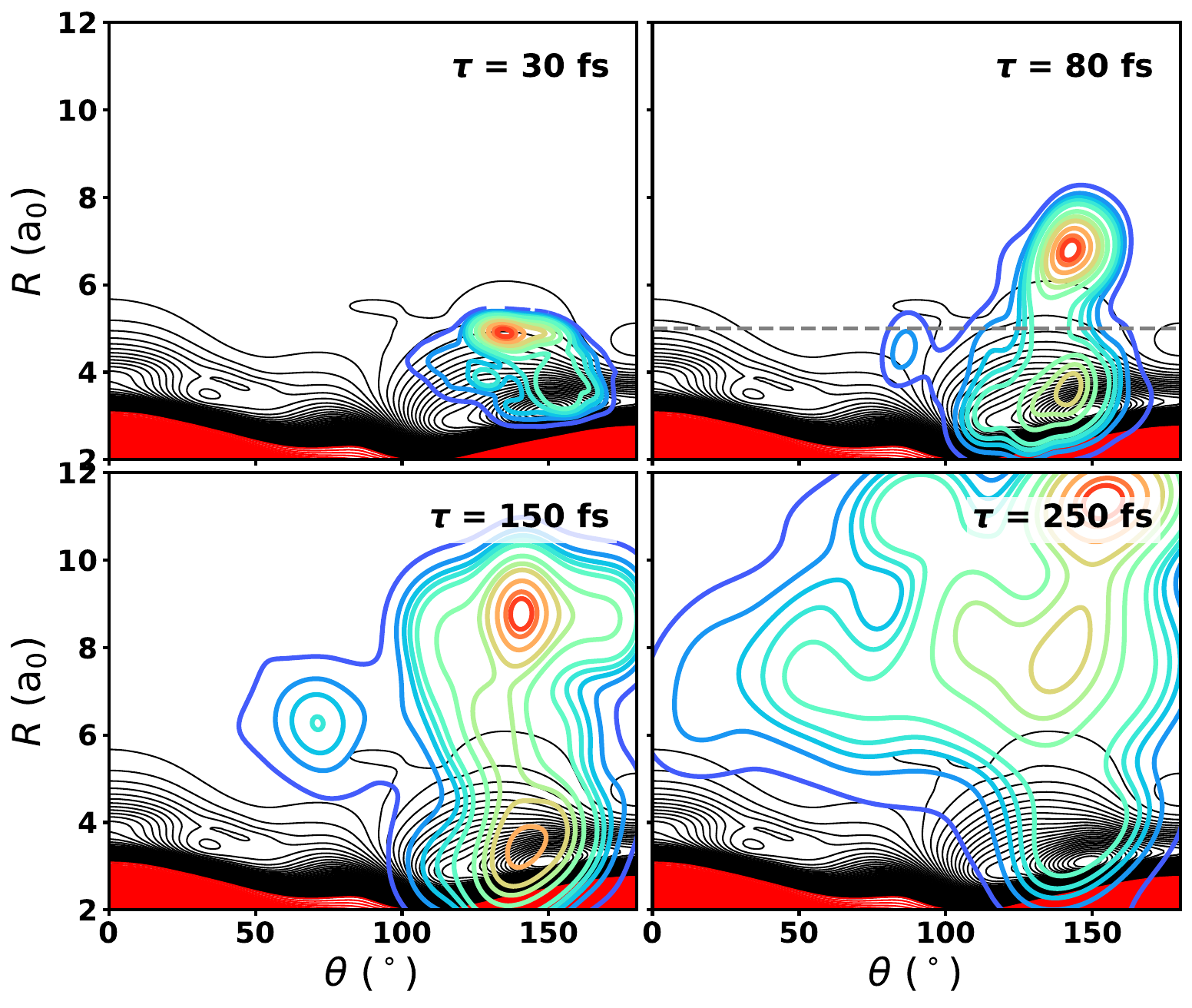}
    \caption{Distributions of $P(R,\theta)$ for "all" trajectories
      with lifetimes $\tau_{\rm life} \leq 1$ ps at different time
      delays, $\tau$, after synchronization. Trajectories are
      synchronized at $R=3.0,a_0$ and obtained using stratified
      sampling of the impact parameter. The distributions are overlaid
      on the potential energy surface (PES)\cite{MM.no2:2020}; black
      and red contours denote negative and positive energies,
      respectively. Colored contours for $P(R,\theta)$ indicate
      1--90\% of the maximum trajectory density (1, 3, 5, 7, 10, 20,
      30, 40, 60, 80, and 90\%) and follow a rainbow color map from
      violet to red. Depending on the value of $\tau$ considered, each
      panel contains $\sim 3000$ trajectories. Panels for $\tau = 80$
      fs and $\tau = 250$ fs indicate that the loose transition state
      is between $R=5$ and 6 a$_0$, indicated by the dashed horizontal
      line.}
    \label{fig:pofrtheta1}
\end{figure}

\noindent
Next, the dependence of the final $P(j')$ distributions on the initial
rotational state $P(j)$ distribution were analyzed. Two different
initial $P(j)$ were considered (see Figure \ref{sifig:pofj-ini}) : the
first involves stratified sampling of the impact parameter $b$ and the
second uses $j-$values drawn from the measured $P(j)$ from
photodissociation experiments, converted to the corresponding impact
parameter $b$ as described in the Methods section.\cite{reisler:1994}
Running 500\,000 independent simulations, the final $P(j')$ was
determined and separately analyzed for O-atom exchange (blue traces)
and inelastic trajectories (yellow), see Figure
\ref{sifig:pofj-reisler}. Both types of initial conditions yield a
featureless $P(j')$ for trajectories in which O-atom exchange occurs,
and $P(j')$ peaks at $10 < j' < 15$. The distribution is somewhat more
peaked around $j' = 12$ using the measured $P(j)$ and extends up to
$j' \sim 25$ compared with $j' \sim 22$ when drawing from a stratified
distribution $P(b)$. Generally, the $P(j')$ for atom-exchange
vs. inelastic trajectories differ little for both types of initial
conditions. Furthermore, the overall shapes of the final state
distributions from the two rather different types of initial $P(j)$
conditions are surprisingly similar, see Figure
\ref{sifig:pofj-reisler}.\\

\begin{figure}[h!]
    \centering
    \includegraphics[width=1.0\linewidth]{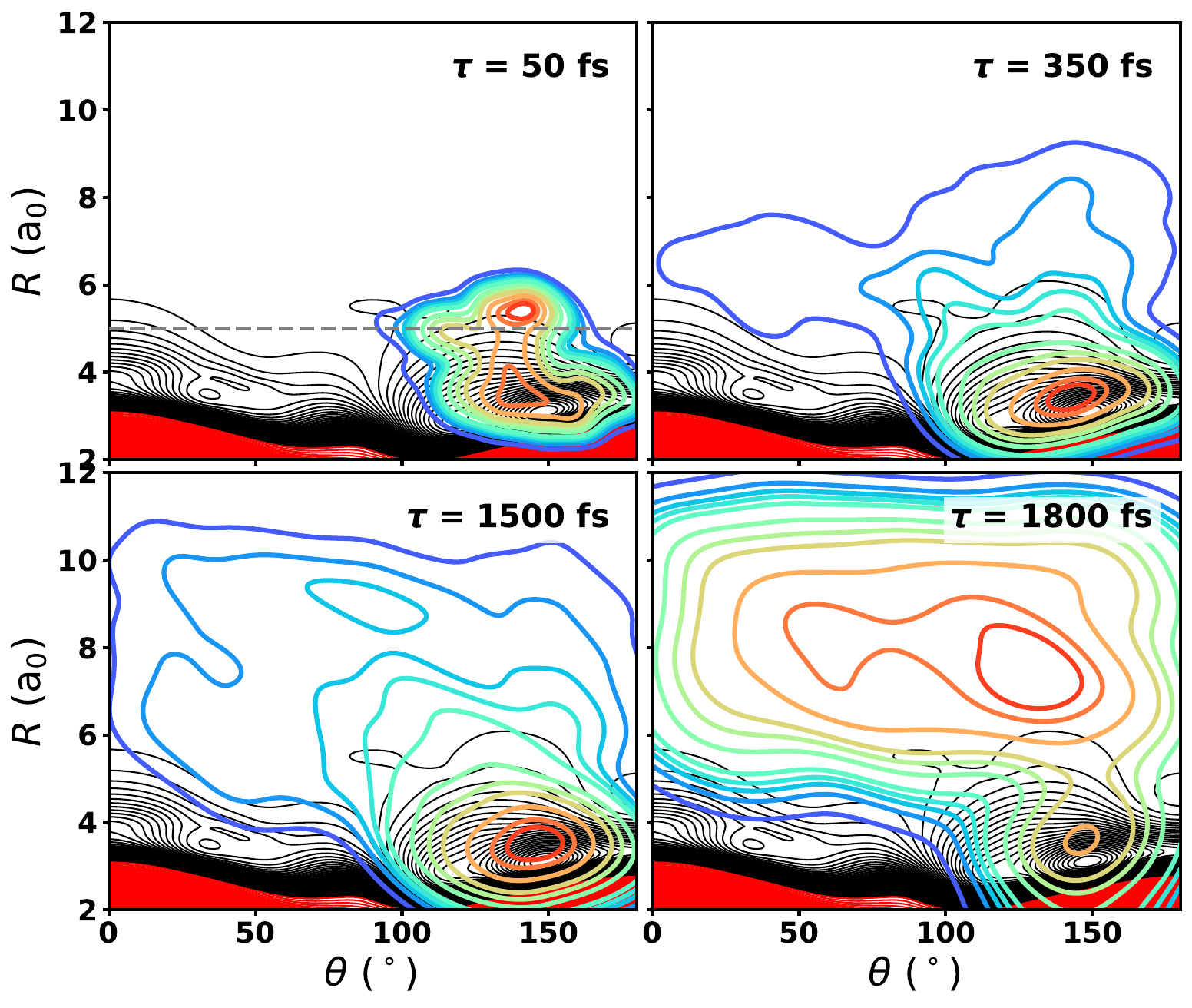}
    \caption{Distributions of $P(R,\theta)$ for "all" trajectories
      with lifetimes $2 \leq \tau_{\rm life} \leq 3$ ps at different
      time delays, $\tau$, after synchronization. Trajectories are
      synchronized at $R=3.0$ a$_0$ and obtained using stratified
      sampling of the impact parameter. The distributions are overlaid
      on the potential energy surface (PES)\cite{MM.no2:2020}; black
      and red contours denote negative and positive energies,
      respectively. Colored contours for $P(R,\theta)$ indicate
      1--90\% of the maximum trajectory density (1, 3, 5, 7, 10, 20,
      30, 40, 60, 80, and 90\%) and follow a rainbow color map from
      violet to red. Depending on the value of $\tau$ considered, each
      panel contains $\sim 2200$ trajectories. The panels for $\tau =
      50$ fs and $\tau = 1800$ fs indicate that the loose transition
      state is between $R=5$ and 6 a$_0$, indicated by the dashed
      horizontal line.}
    \label{fig:pofrtheta2}
\end{figure}

\noindent
Finally, the conformational space sampled by the synchronized
trajectories was determined as a function of time $\tau$ after
synchronization, with all trajectories starting from the initial "time
zero" complex formation. Here, the spatial distribution function
$P(R,\theta)$ was computed for trajectories with different lifetimes,
see Figures \ref{fig:pofrtheta1} and \ref{fig:pofrtheta2}. In order to
generate the cumulative spatial distributions $P(R,\theta)$, the
trajectories used for analysis were first synchronized, see Methods
section and Figure \ref{fig:sketch}. Figure \ref{fig:pofrtheta1} shows
$P(R,\theta)$ for $\sim 3000$ trajectories with lifetimes $\tau_{\rm
  life} < 1$ ps, at delay times of 30, 80, 150 and 250 fs after
synchronization. The underlying 2-dimensional rendering of the PES
$V(R,\theta; r = 2.26 {\rm a}_0)$ provides a reference to relate the
time evolution of $P(R,\theta)$ with features of the
PES.\cite{MM.no2:2020} The global minimum (see Supporting
Information\cite{MM.no2:2020}) is at an ONO-angle of $134.2^\circ$
with a secondary minimum at $\theta = 40.6^\circ$.\\

\noindent
At 30 fs after formation of hot ground state [NO$_2$]$^\nmid$, the
density predominantly samples the deep minimum corresponding to the
NO$_2$ equilibrium structure. After 80 fs, parts of the distribution
separate and sample either regions with $\theta_{\rm ONO} \sim
90^\circ$ or $R \sim 6$ a$_0$, i.e. well away from the global
minimum. At the next time point - 150 fs after synchronization - the
two separated parts keep moving away from the global minimum, with
some density already reaching $R \sim 10$ a$_0$, i.e. in the
dissociative region. After 250 fs, the entire $\theta-$range is
populated and the maximum of a second part of the distribution reaches
$R \sim 8$ a$_0$ and continues moving towards dissociation. \\

\noindent
Figure \ref{fig:pofrtheta2} shows $P(R,\theta)$ for $\approx 2200$
longer-lived trajectories ($2 \leq \tau_{\rm life} \leq 3$ ps). Here,
at 50~fs after [NO$_2$]$^\nmid$ formation, $P(R,\theta)$ has already
separated into two distinct distributions. One part continues to
sample the deep minimum corresponding to the equilibrium structure of
NO$_2$, whereas the other is centered around $R \sim 5$~a$_0$. By
350~fs after synchronization, the latter distribution started to move
toward dissociation, with part of the density reaching $R \sim
8$~a$_0$ and shifting to the second minimum of the PES. The remaining
density continues to broaden around the global minimum. By 1500~fs
after synchronization, the maximum ($28$ \%) of the density is
centered around global minimum and $14$ \% of the trajectories
progressed towards dissociation. By 1800~fs, the dissociating
distribution has propagated further towards larger values of $R$
($\sim 64$ \% of the trajectories have reached $R \sim 7$~a$_0$ and
longer), whereas 7\% of the density remains localized near the global
minimum and an increasing fraction of the trajectories having
progressed toward complete dissociation.  \\

\subsection{Coulomb Explosion Imaging}
For direct comparison with the TR-CEI measurements,\cite{stolow:2019}
the NO$^+$ and O$^+$ ion pair Coulomb recoil energy (CRE) as a
function of time was computed. In CEI, it is assumed
\cite{stapelfeldt1998time} that the distance between recoiling neutral
fragments (in a molecular dissociation) is preserved during SFI, in
this case generating the doubly charged cation. The impulsively
prepared NO$^+$ and O$^+$ ion pair then dissociates with a CRE
determined only by the scalar distance between them. Thus, the
relevant progression coordinate $R$ is the scalar distance between the
neutral ground state NO (center-of-mass) and the O photofragment. The
time evolution of their ground state separation thus determines the
time evolution of the ion pair CRE: the recoil energy falls inversely
with the scalar distance between the separating neutral ground state
photofragments.\\

\noindent
Figure \ref{sifig:pofr-time} compares the evolution of $P(R)$ after
synchronization for exchange (left column) and inelastic (right
column) NO$_2$ dissociation trajectories, grouped according to their
dissociation lifetimes: 0.25--0.5 ps (top row), 2--3 ps (middle row),
and 10--15 ps (bottom row). In each panel, the black curve represents
the potential energy\cite{MM.no2:2020} along the N--O separation
coordinate $R$. The colored curves correspond to vertically offset
snapshots of $P(R)$ at different times following synchronization
$\tau$: $1$, $5$, $25$, and $50$ fs, followed by $25\%$, $50\%$,
$75\%$, $90\%$, and $100\%$ of the corresponding channel lifetime. For
the longest-lifetime trajectories (panels E and F), an additional
curve at $95\%$ of the lifetime is included. The color gradient from
dark purple to light yellow represents increasing time delays $\tau$
following synchronization.\\

\noindent
In all cases, $P(R)$ rapidly re-synchronizes after a brief initial
transient and remains largely trapped within the well before
dissociation. With increasing $\tau$, the probability density
progressively shifts toward larger internuclear separations,
reflecting the evolution towards dissociation. For the shortest
lifetimes (panels A and B), the wavepacket leaves the well quickly and
remains relatively compact as it moves outward. For intermediate
lifetimes (panels C and D), dissociation proceeds more slowly and the
probability distributions exhibit increased broadening before reaching
larger separations. In the longest-lifetime cases (panels E and F),
the distribution remains trapped near the potential well for a
substantially longer time and becomes strongly delocalized at later
stages, spreading over a wide range of $R$ values (approximately
$5$--$17\,a_0$) before complete dissociation. Exchange and inelastic
channels show comparable qualitative behaviour for the two longer
lifetimes, whereas for lifetimes in the 0.25--0.5 ps interval (panels
A/B) they clearly differ. This suggests that the underlying
phase-space dynamics for short-lived [NO$_2$]$^\nmid$ differs between
these two channels.\\

\noindent
Figure \ref{sifig:pofr-area-1} shows the time evolution of the
integrated densities of $P(R)$ for $0 \leq R \leq 5$ a$_0$ and for $5
< R \leq \infty$, separating the bound well from the dissociation
continuum, for three time windows (0.25--0.5, 2--3, and 10--15 ps) and
two collision mechanisms (exchange, left column; inelastic, right
column). By construction, the trapping time increases with the
lifetime of the hot [NO$_2$]$^\nmid$. However, once dissociation
begins the transition is rapid, producing a sharp transfer of
population to the continuum rather than a gradual decay, see also
Figure \ref{sifig:pofr-area}. Overall, longer hot [NO$_2$]$^\nmid$
lifetimes correspond to longer-lived trapped states, while exchange
and inelastic mechanisms exhibit similar behavior, with only minor
differences in the onset of dissociation, except for the shortest
lifetime considered.\\

\begin{figure}[h!]
    \centering
    \includegraphics[width=1.0\linewidth]{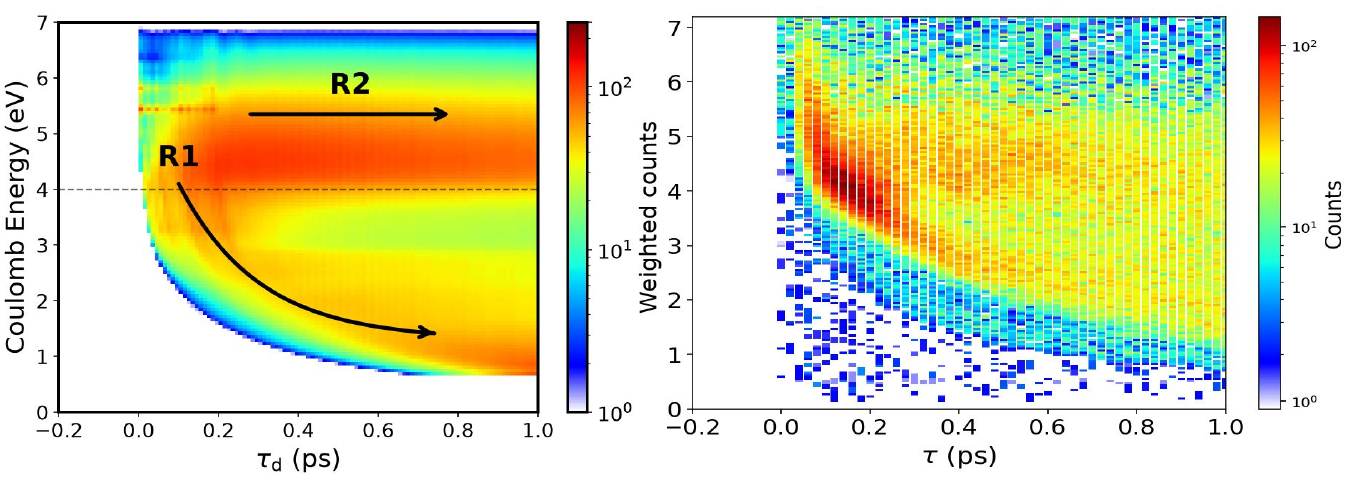}
    \caption{Coulomb repulsion energy between NO$^{+}$ and O$^{+}$ as
      a function of the time $\tau_{\rm d}$ (ps) after synchronization
      for 216\,249 trajectories forming the NO$_2$ complex (out of
      500\,000 runs in total). Trajectories were synchronized by
      determining the first instance at which the distance $R$ between
      the center of mass of NO$^{+}$ and O$^{+}$ reached $3.0\,a_{0}$,
      which corresponds to $\tau = 0$. Simulations used stratified
      sampling of the impact parameter $b$. Panels A and B compare the
      results of the present simulations with the TR-CEI experimental
      data from Ref.\cite{stolow:2019}. Panel A was constructed using
      a superposition of time-shifted copies of the scaled trajectory
      histogram, \(M\) (raw counts/250), given by \(0.75\,M(\tau_d) +
      0.20\,M(\tau_d + 100\,\mathrm{fs}) + 0.05\,M(\tau_d +
      200\,\mathrm{fs})\) using weights from previous wavepacket
      simulations.\cite{arasaki:2007} The delayed terms are shifted
      only along the time axis (\(\tau_d\)), while the Coulomb
      repulsion energies were scaled to 60\% of their original values
      (i.e., reduced by 40\%) to place them on a scale comparable to
      the measurements. Reduction of the Coulomb energy is equivalent
      to "shielding". Figure \ref{sifig:ecoul} reports computed TR-CEI
      spectra for exchange, inelastic, and elastic processes.}
    \label{fig:ecoul}
\end{figure}

\noindent
For each time delay after [NO$_2$]$^\nmid$ formation - defined as the
time at which $R = 3.0$ a$_0$ for the first time - the Coulomb energy
between NO$^+$ and O$^+$ was calculated using $P(R)$. This yielded a
2d-map relating the delay time $\tau_{\rm d}$ after NO$_2$ formation
and the Coulombic repulsion energy. Figure \ref{fig:ecoul}A shows the
computed result which is compared with the TR-CEI measurements, Figure
\ref{fig:ecoul}B.\cite{stolow:2019} The computed and measured maps are
qualitatively similar, with both showing two distinct regimes. Regime
R1 corresponds to a rapidly decaying population for which the CRE
decreases on the sub-ps time scale towards almost zero,
i.e. dissociation into well-separated neutral NO+O fragments for which
the CEI energy becomes negligible. In contrast, regime R2 features, on
the ps time scale, an approximately constant CRE of $\sim 5$ eV. It
should be noted that in the computations, the Coulomb energy was
scaled down by 40 \% to be comparable with the measurements. This is
not unexpected due to screening of the point Coulomb potential due the
remaining electron density between the separating
cations.\cite{xu2025ultrafast} On the $\sim 1$ ps timescale, the
recoiling NO$^+$ and O$^+$ fragments may not yet be well described as
separated point charges. Although the interaction should approach the
bare Coulomb repulsion between two singly charged fragments at long
range, the early-time dynamics can involve a softened effective
repulsion due to charge delocalization, residual electron density
between the fragments, and the finite spatial extent of the NO$^+$
fragment. Thus, apparent screening of the Coulomb interaction can play
a role during the first picosecond of separation. \\

\begin{figure}
    \centering
    \includegraphics[width=1.0\linewidth]{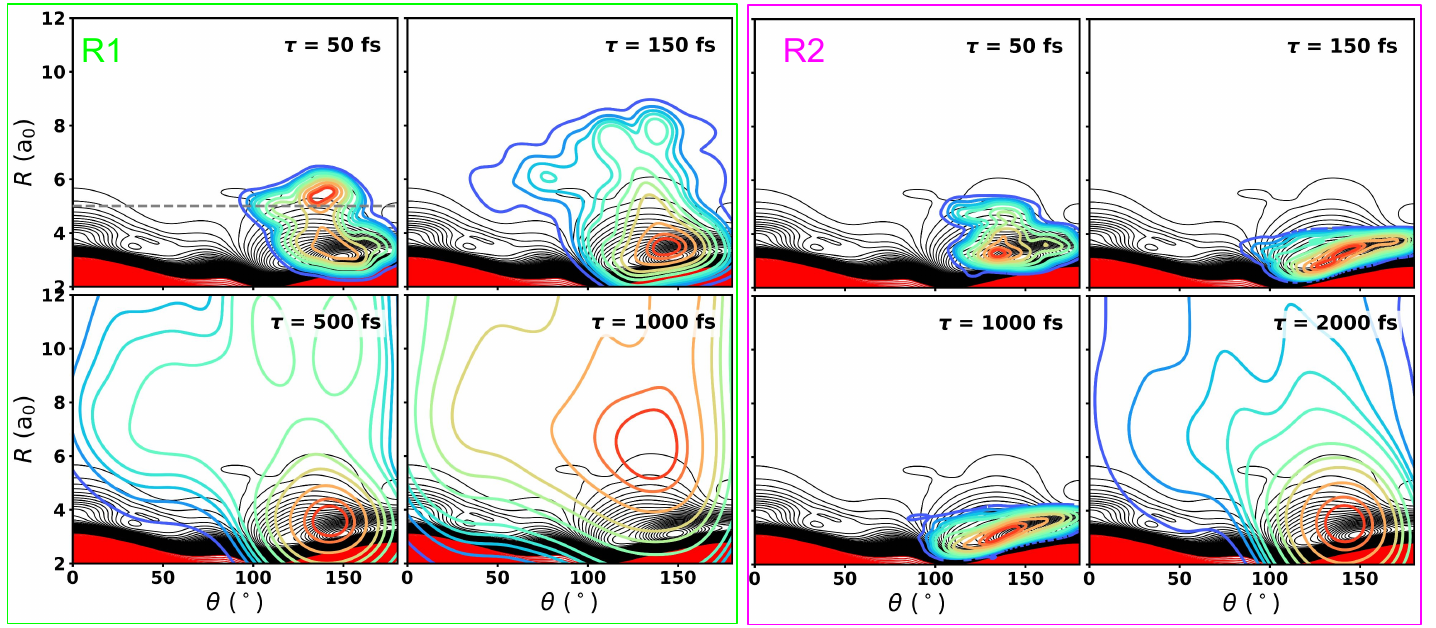}
    \caption{Distributions of $P(R,\theta)$ for trajectories
      contributing to the two regimes, R1 and R2, in Figure
      \ref{fig:ecoul}. In the R1 regime, the Coulomb energy decays on
      a timescale of approximately 1 ps, whereas for regime R2 it
      remains approximately constant at $E_{\rm coul} \sim 4$ eV
      during the first 1 ps. The distributions were generated from
      3000 trajectories randomly selected from all trajectories
      contributing to R1 and R2, respectively.}
    \label{fig:ecoul_R1-R2}
\end{figure}

\noindent
Figure \ref{fig:ecoul_R1-R2} shows the dependence of the 2-dimensional
probability distributions $P(R,\theta)$ on the time delay after
synchronization $\tau$ for trajectories contributing to R1 and R2,
respectively. In R1, the distribution rapidly moves away from the
minimum of the PES and within 1 ps the maximum of the distribution has
reached the loose transition state $q_{\rm TS}$, see Figure
\ref{fig:sketch}. In contrast, for R2 $P(R,\theta)$ remains trapped
above the minimum energy geometry of the PES and only starts to move
away on the 2 ps time scale. This can also be seen in the lifetime
distribution of the trajectories contributing to R1 and R2, shown in
Figure \ref{sifig:life-coul}.

\subsection{Phase Space Analysis}
After establishing that the QCT simulations provide a qualitatively
correct picture of the TR-CEI experiments, the underlying reaction
dynamics were analyzed in more detail. A prime question concerns the
statisticality of the dynamics, including energy redistribution,
randomization of the dynamics, phase space sampling, ergodicity,
barrier recrossing, and related microscopic behavior. For a phase
space view of the dissociation process, it is useful to generate
Poincar\'{e} maps which relate position and momentum (velocity) of the
particles. In the following, three different aspects are analyzed in
more detail: the fraction of available phase space sampled by a
typical trajectory with a given lifetime, the distribution of
residence and recurrence times, and the dependence of Lyapunov
exponents on the lifetimes of neighbouring trajectories.\\

\begin{figure}[h!]
\includegraphics[width=0.8\textwidth]{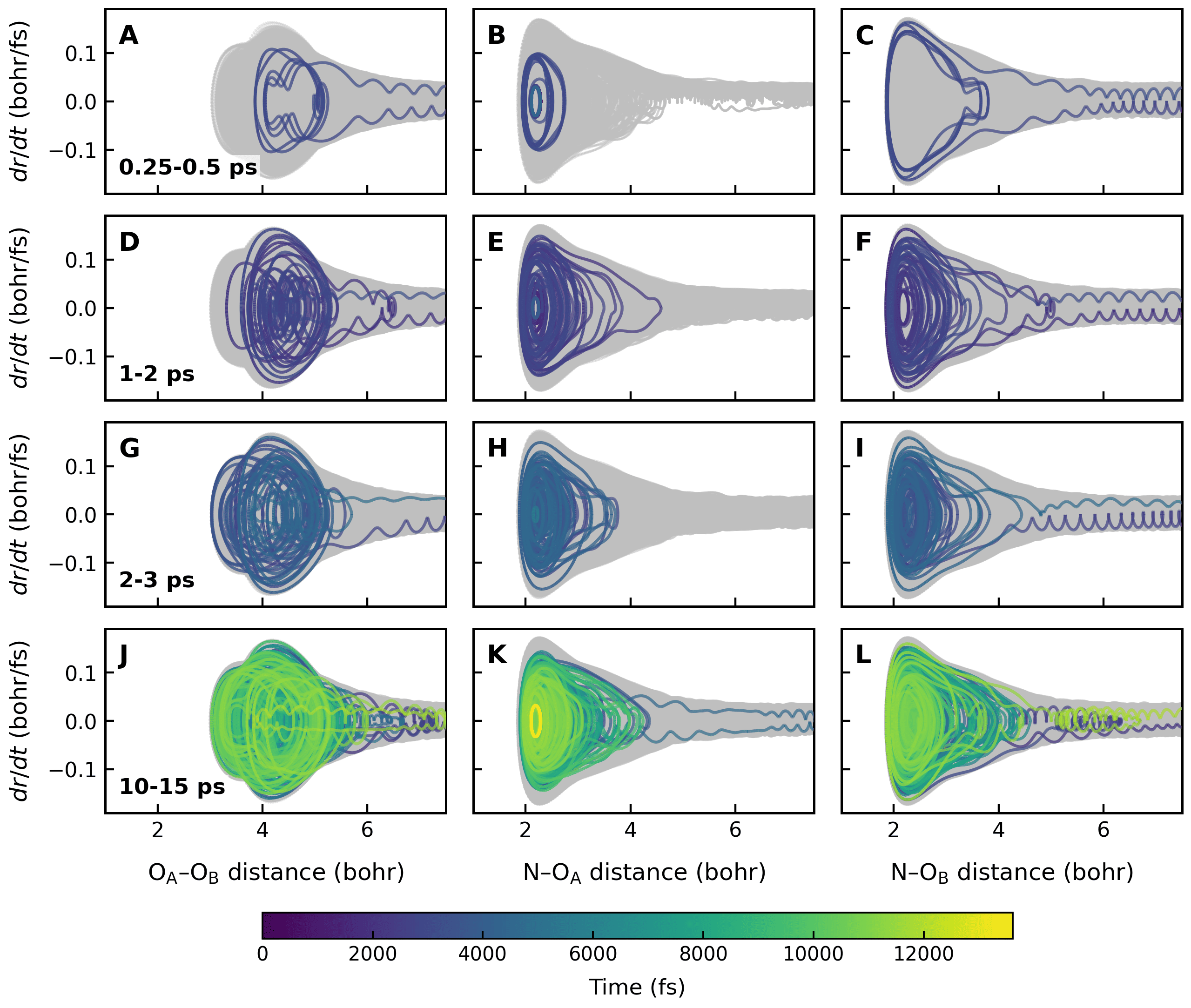}
\caption{Phase space trajectories for selected lifetimes of the
  [NO$_2$]$^\nmid$ collision complex. The grey areas were generated
  from all trajectories with given lifetimes, whereas the colored
  trajectory is one representative inelastic phase space trajectory
  with given lifetime. Each row corresponds to a different lifetime
  interval: 0.25–0.5 ps, 1–2 ps, 2–3 ps and 10–15 ps. The three
  columns report the phase-space projections for the
  O$_{\mathrm{A}}$–O$_{\mathrm{B}}$ (left), N–O$_{\mathrm{A}}$
  (middle), and N–O$_{\mathrm{B}}$ (right) distances,
  respectively. They are generated from the internuclear separation
  and the corresponding radial velocity $dr/dt$.}
\label{fig:phase_space}
\end{figure}

\noindent
{\it Analysis of Fractional Area Sampled:} As the highly energized,
``hot'' parent species [NO$_2$]$^\nmid$ is generated from a
half-collision NO+O at low collision energy, it is reasonable to
assume that the ensemble of longest-lived trajectories sampling the
NO$_2$ well provide a meaningful rendering of the total available
phase space. Such phase space trajectories are shown in Figure
\ref{fig:phase_space}. Trajectories were grouped according to their
lifetime. The four rows in Figure~\ref{fig:phase_space} correspond to
the lifetime intervals $0.25 \leq \tau_{\rm life} \leq 0.5$~ps, $1
\leq \tau_{\rm life} \leq 2$~ps, $2 \leq \tau_{\rm life} \leq 3$~ps,
and $10 \leq \tau_{\rm life} \leq 15$~ps, respectively. For each
interval, phase space data from a large ensemble of independent
trajectories were accumulated to construct the gray background
distributions. Superimposed on each distribution is one representative
inelastic scattering trajectory from the corresponding lifetime
interval, from which the fractional area explored by an individual
trajectory can be determined. As the lifetime of the highly excited
collision complex increases (large $\tau$), the fractional area
covered by a single trajectory progressively increases. A
complementary phase space representation based on the Jacobi
coordinate, $R$, and its conjugate velocity, $dR/dt$, is shown in
Figure~\ref{fig:si_phase_R} for representative trajectories from the
same lifetime intervals.\\

\noindent
An alternative representation based on the coordinate
$\Delta=r_{\mathrm{N-O_A}}-r_{\mathrm{N-O_B}}$ further illustrates how
the dissociation dynamics changes with the lifetime of the complex
(Figures~\ref{fig:si_phase_delta} and ~\ref{fig:si_delta_time}).  For
short-lived trajectories, $\Delta$ rapidly departs from zero
(i.e. equal N-O bond lengths) and adopts either a positive or negative
value, indicating an early selection of the final dissociation
arrangement. As the lifetime increases, the trajectories undergo
increasingly prolonged oscillations around $\Delta=0$ before
separation. Correspondingly, the $(\Delta,d\Delta/dt)$ projections
evolve from a comparatively small number of loops into densely wound
trajectories that repeatedly cross both $\Delta=0$ and
$d\Delta/dt=0$. These repeated crossings reflect large amplitude
motion between configurations in which the two N--O distances are
comparable and indicate that the system progressively loses memory of
its initial arrangement before dissociation. This behavior is
consistent with the broader phase-space sampling observed for the
long-lived trajectories and with the gradual approach to statistical
dynamics.\\

\begin{figure}[h!]
\includegraphics[width=0.6\textwidth]{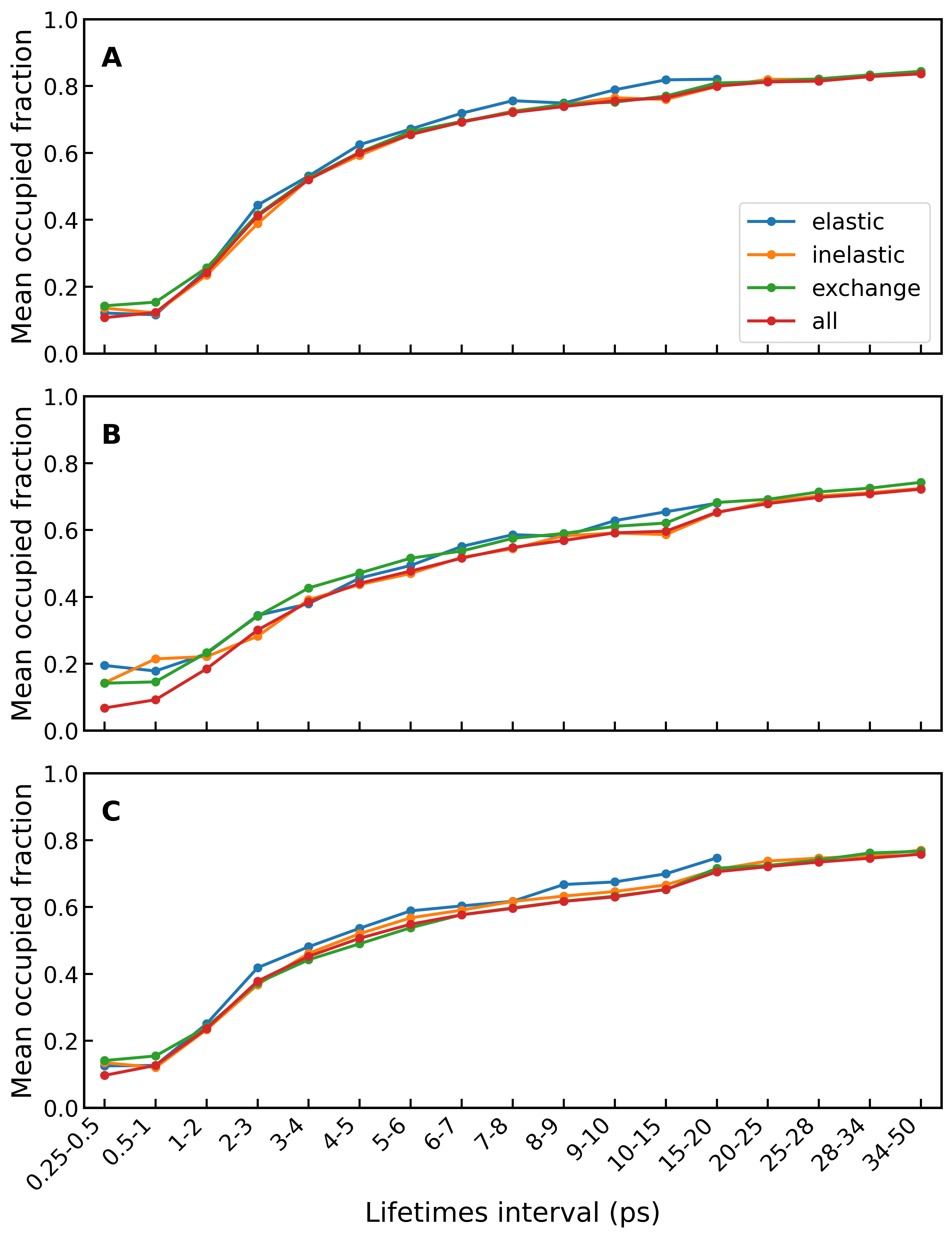}
\caption{Mean occupied phase-space area fraction, $\langle
  A_i/A_{\mathrm{total}} \rangle$, as a function of the lifetime
  interval for the O$_{\mathrm{A}}$--O$_{\mathrm{B}}$ coordinate in
  panel A, the N--O$_{\mathrm{A}}$ coordinate in panel B, and the
  N--O$_{\mathrm{B}}$ coordinate in panel C.  Results are shown
  separately for elastic, inelastic, and exchange trajectories,
  together with the average over all trajectories. For elastic
  trajectories, the statistics are insufficient for $\tau_{\rm life} >
  15$~ps, and the corresponding data are therefore not reported.  }
\label{fig:frac_area}
\end{figure}

\noindent
As shown in Figure \ref{fig:frac_area}, the quantity $\langle
A_i/A_{\mathrm{total}} \rangle$ measures the mean fraction of the
total accessible phase space area explored by individual trajectories
within each lifetime interval. The occupied fraction was calculated
from 100 representative trajectories for each process and normalized
to the total accessible region in the corresponding phase-space
projection. For elastic trajectories, the statistics is insufficient
for $\tau_{\rm life} > 15$ ps and these data are not reported. The
mean occupied phase space fraction increases with $\tau_{\rm life}$
for the three internal-coordinate projections. Evidently, long-lived
trajectories explore larger portions of the accessible phase space
than short-lived ones: 80 \% vs. $\sim 40$ \%. The convergence of the
occupied fractions for longer lifetimes suggests that the dynamics
approaches a more space-filling regime, where trajectories have
sufficient time to visit a larger fraction of the available
configurations before leaving the interaction region.\\

\noindent
{\it Analysis of Recurrence Times:} A complementary analysis is that
of recurrence times, $\Delta^{\rm rec}$, see Figure
\ref{fig:sketch}. The recurrence time for a single trajectory is
defined as the time interval which a particular degree of freedom
spends between successive crossings of a given target value. Depending
on the lifetime of the energized [NO$_2$]$^\nmid$ complex, multiple
recurrence (and residence) times occur for a single trajectory. The
set of recurrence times summed over all trajectories with a given
lifetime (or lifetime interval) is the distribution of recurrence
times. Examples are given in Figure \ref{sifig:distributions}.\\

\noindent
For each lifetime interval, all recurrence (or residence) events
extracted from the trajectory ensemble were pooled into a single
dataset and represented as histograms. To characterize the short-time
behavior of the distributions while minimizing the influence of the
long-time tail, the fitting was restricted to $0 < \Delta^{\rm rec} <
250$ fs and $0 < \Delta^{\rm res} < 800$ fs, respectively. The
distributions were fitted by maximum-likelihood estimation using the
Gamma probability density function
\begin{equation}
f(\tau;\alpha,\theta)= \frac{\tau^{\alpha-1}e^{-\tau/\theta}}
{\Gamma(\alpha)\theta^{\alpha}},
\end{equation}
where $\alpha$ is the shape parameter, $\theta$ is the scale
parameter, and $\lambda=1/\theta$ is the corresponding rate. To allow
a direct comparison with the histogram counts, the fitted probability
density was rescaled by the number of events included in the fit and
by the histogram bin width. The resulting curves therefore represent
the expected number of events per histogram bin rather than a
normalized probability density.\\

\noindent
The recurrence- and residence-time distributions in
Figure~\ref{sifig:distributions} reveal two complementary aspects of
the phase space dynamics. The recurrence-time distributions, shown in
Figures~\ref{sifig:distributions}A/C/E/G quantify the time intervals
between successive returns of a trajectory to a prescribed region or
dividing surface in phase space, analogous to the $\tau_{ij}$
intervals illustrated in Figure~\ref{fig:sketch}. These distributions
are strongly skewed toward short times and are well described by
gamma-like fits, indicating that once a trajectory enters the
interaction region, repeated returns occur rapidly. The fitted shape
parameters $\alpha \sim 1$ further suggest an approximately memoryless
return process, consistent with efficient local mixing inside the
transient [NO$_2$]$^\nmid$ collision complex.\\

\noindent
In contrast, the residence-time distributions in
Figures~\ref{sifig:distributions}B/D/F/H describe the duration for
which trajectories remain in the NO$_2$ well before an attempt to
dissociate is made. These distributions are broader and characterized
by larger $\alpha-$values than the recurrence-time distributions. The
similarity of the distributions across increasing lifetime intervals
in Figure~\ref{sifig:distributions} suggests that long-lived
trajectories are not necessarily associated with a qualitatively
different local return mechanism; rather, their longer lifetimes allow
a larger number of recurrence and residence events before final
dissociation or separation.\\

\noindent
The Gamma shape parameters obtained from the recurrence-time and
residence-time distributions exhibit a dependence on the lifetime
interval (Figure~\ref{fig:combi_r_and_r}). The recurrence-time
parameter, $\alpha_{\mathrm{rec}}$, increases for short-lived
complexes and reaches an approximately constant value for $\tau
\gtrsim 2$--$3$~ps. By contrast, the normalized residence-time
parameter, $\alpha_{\mathrm{res}}^{\mathrm{norm}}$, reaches a maximum
within this lifetime range and subsequently decreases for longer-lived
complexes.  Interestingly, this crossover occurs on a timescale
similar to that at which the relative populations of exchange and
inelastic trajectories become approximately stable, as shown in
Figure~\ref{fig:poftlife}.\\

\noindent
{\it Lyapunov Exponents:} The Lyapunov exponents, a measure of the
divergence of neighbouring trajectories, were analyzed for
trajectories exhibiting different lifetimes. It should be noted that
Lyapunov exponents provide a measure of dynamical instability and
local chaoticity. Lyapunov exponents are not direct measures of the
``statisticality'' of a process. Rather, they provide a complementary
description of the dynamical instability and evolution of the
collision complex. To characterize this instability within the
interaction region, $\langle\lambda_{\mathrm{life}}\rangle$ was
defined as the mean Lyapunov exponent evaluated over the lifetime of
the complex for all trajectories belonging to a given lifetime
interval. As shown in Figure~\ref{sifig:lyapunov_coef_coll},
$\langle\lambda_{\mathrm{life}}\rangle$ increases rapidly for the
shortest-lived trajectories, from approximately 0.027 fs$^{-1}$ in the
$0$--$0.25$~ps interval to $\sim 0.040$~fs$^{-1}$ for lifetimes of
2--3~ps. For longer-lived trajectories,
$\langle\lambda_{\mathrm{life}}\rangle$ fluctuates in the interval
$[0.041,0.043]$ fs$^{-1}$. Correspondingly, the characteristic
randomization time, $\tau_{\mathrm{rand}}^{\mathrm{life}}
=1/\langle\lambda_{\mathrm{life}}\rangle$, decreases from $\sim 37$~fs
for the shortest-lived trajectories to $\sim 25$~fs at 2--3~ps, after
which it remains within a relatively narrow range of approximately
$23$--$25$~fs. This behavior indicates that dynamical instability
develops rapidly during the first few picoseconds, whereas
longer-lived complexes exhibit a broadly similar characteristic
randomization timescale. Interestingly, this transition occurs within
approximately the same lifetime range highlighted by the dashed line
in the inset of Figure~\ref{fig:poftlife}, beyond which the relative
fractions of inelastic and exchange trajectories approach comparable
values and remain close to $50\%$. The similar timescales suggest the
emergence of a common long-lived dynamical regime beyond approximately
2--3~ps.\\

\noindent
The time evolution of the separation between initially neighboring
trajectories provides a direct illustration of the dynamical
instability quantified by the Lyapunov exponent
(Figure~\ref{fig:si_distance}). For both representative trajectories,
the quantity $\ln[D(t)/D(0)]$ increases predominantly within the
shaded lifetime interval, confirming that the strongest divergence
develops when the system samples the interaction region. For the
short-lived trajectory, the separation grows progressively throughout
most of its lifetime, whereas the longer-lived trajectory exhibits a
rapid initial increase followed by saturation well before
dissociation. This saturation does not imply the disappearance of
chaotic dynamics; rather, it reflects the finite extent of the
available phase space region, limiting the accessible separation
between the trajectories. These examples therefore support the use of
the renormalized Lyapunov analysis to characterize local instability
and indicate that most of the dynamical divergence is established
during the early part of the complex lifetime.\\

\begin{figure}[h!]
	\includegraphics[width=0.9\textwidth]{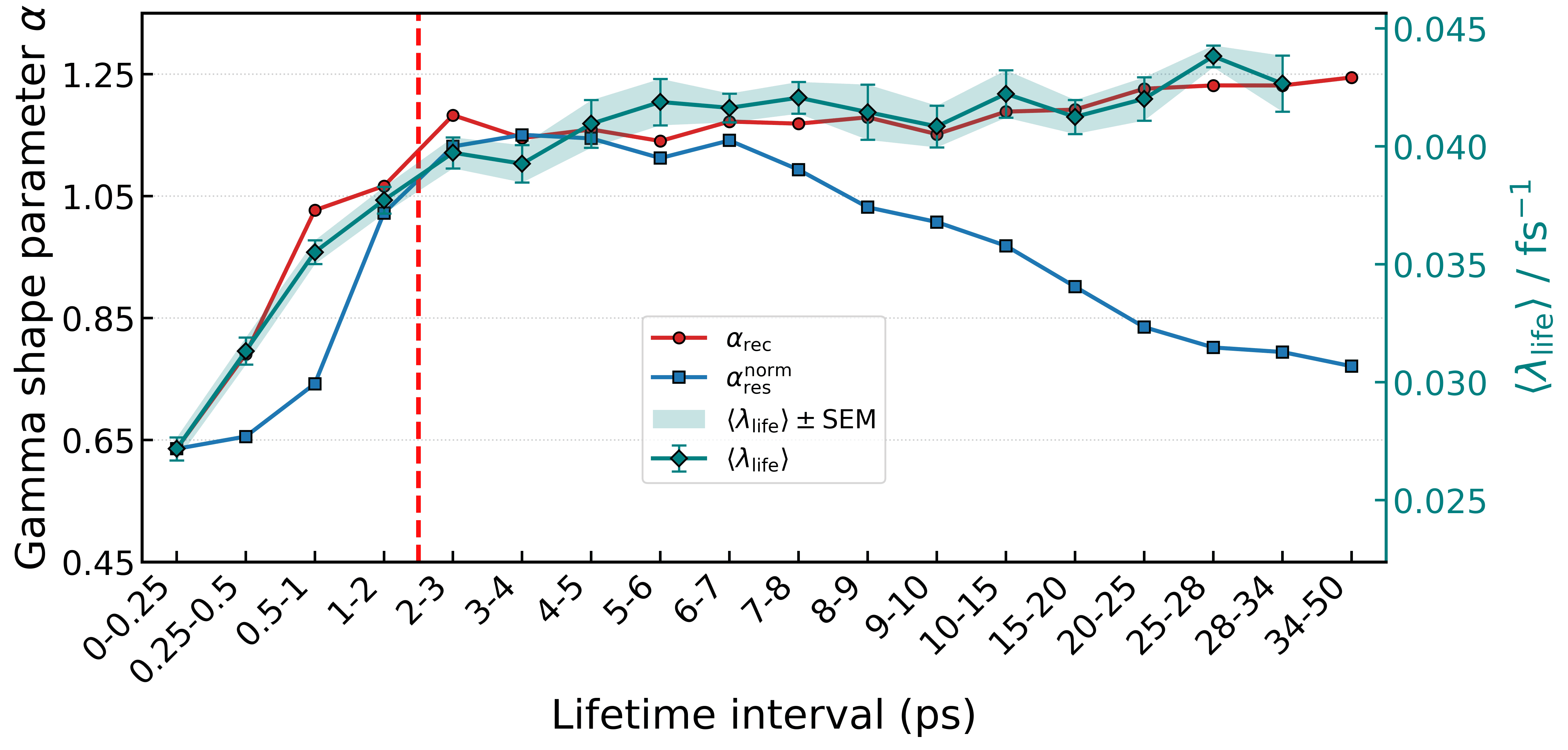}
\caption{Gamma-distribution shape parameters and Lyapunov exponent as
  functions of the $\mathrm{NO_2}$ lifetime interval. The
  recurrence-time shape parameter, $\alpha_{\mathrm{rec}}$ (red
  circles), and the normalized residence-time shape parameter,
  $\alpha_{\mathrm{res}}^{\mathrm{norm}}$ (blue squares), are shown on
  the left axis. The mean lifetime Lyapunov exponent,
  $\langle\lambda_{\mathrm{life}}\rangle$ (teal diamonds), is shown on
  the right axis; error bars and the shaded region represent the
  standard error of the mean (SEM). The residence-time values were
  normalized to match $\alpha_{\mathrm{rec}}$ in the first lifetime
  interval. The vertical red dashed line marks the boundary at
  $\tau_{\mathrm{life}}=2$~ps.}
\label{fig:lyapunov_coef_coll}
\end{figure}

\noindent
Figure~\ref{fig:lyapunov_coef_coll} combines the recurrence- and
residence-time statistics with the mean Lyapunov exponent to
characterize the progressive randomization of the dynamics. For
$\tau_{\mathrm{life}}<2$~ps, $\alpha_{\mathrm{rec}}$ increases
rapidly, indicating that successive returns become less broadly
distributed as the complex explores a larger region of phase
space. Over the same range, $\alpha_{\mathrm{res}}^{\mathrm{norm}}$
and $\langle\lambda_{\mathrm{life}}\rangle$ also change markedly,
showing that the local trapping dynamics and the sensitivity to
initial conditions evolve together. Beyond approximately 2-3~ps,
$\alpha_{\mathrm{rec}}$ and the Lyapunov exponents vary more
gradually, whereas $\alpha_{\mathrm{res}}^{\mathrm{norm}}$ reaches a
maximum and subsequently decreases. Thus, the dashed line at 2 ps,
corresponding to the same lifetime boundary marked in
Figure~\ref{fig:poftlife}, indicating the onset of a regime in which
phase space randomization is substantially developed and the dynamical
observables begin to approach more slowly varying dynamical
behavior.\\

\section{Discussion and Conclusion}
The present work discussed the dissociation dynamics of energized
[NO$_2$]$^\nmid$ in the context of recent TR-CEI experiments and
analyzed the phase space dynamics of the unimolecular decomposition
process. Using a validated reactive PES based on an RKHS
representation provides a realistic picture of evolution along the
dissociation $R-$coordinate, as evidenced by the strong agreement
between experiment and theory in Figure \ref{fig:ecoul}. The two
prominent features in the measured $(E_{\rm coul},\tau)-$map, namely
the rapidly decaying (R1) and the quasi-stationary (R2) component,
were successfully captured through dynamics simulations which started
from a slow NO+O collision and the synchronized formation of
[NO$_2$]$^\nmid$. Further analysis associated R1 and R2 with
trajectories having considerably different lifetime distributions in
the [NO$_2$]$^\nmid$ state before breakup.\\

\noindent
For trajectories with lifetimes $\tau_{\rm life} \lesssim 1$ ps, the
probability for inelastic collisions between NO and O differs from
that of O atom exchange. In other words, [NO$_2$]$^\nmid$ lifetimes of
$\sim 1$ ps appear to be too short to fully randomize the internal
energy required to reach the statistically expected 50 / 50 split
between inelastic and reactive collisions, see Figure
\ref{fig:poftlife}. This observation is consistent with convergence of
the Gamma shape parameter $\alpha$ obtained from analysis of the
recurrence time distributions, for which randomization times of 2 to 3
ps are inferred. Finally, analysis of the Lyapunov exponents indicated
that randomization times are 1 ps and longer. Hence, trajectories
corresponding to R1 in Figure \ref{fig:ecoul} are associated with
non-statistical breakup dynamics whereas those in R2 support the
statistical dynamics assumption.\\

\noindent
In the present work the internally ``hot'' ground state triatomic
[NO$_2$]$^\nmid$ was generated from a slow collision between NO$_{\rm
  A}$ + O$_{\rm B}$ under imposed energy and angular momentum
constraints. Synchronization of the trajectories provided a
well-defined ``time zero'' for the unimolecular decay dynamics of
[NO$_2$]$^\nmid$, thereby enabling the comparison of TR-CEI experiment
with theory. From the trajectory data, a unique time-resolved
observable was constructed: the Coulombic Recoil Energy upon SFI. In
contrast, earlier work\cite{arasaki:2007} projected an initial
wavepacket from the ground state onto the electronically excited state
and propagated it through the conical intersection. After reaching
this crossing, parts of the wavepacket switched to the ground state
whereas the remainder continued on the excited state PES until most of
the population was transferred to the ground state after $\sim 200$
fs. From here, dissociation into NO+O was followed in a fashion
analogous to the present work. Differences between the present work
and earlier simulations concern non-adiabatic effects - not included
here since the focus is on [NO$_2$]$^\nmid$ ground state dynamics
after all non-adiabatic processes are essentially complete - and the
higher quality of the underlying PES(s) in the present work and the
construction of new time-resolved observables.\\

\noindent
Time-Resolved Coulomb Explosion Imaging (TR-CEI) is a novel
experimental probe of ground state unimolecular dissociation
dynamics. It provides a direct map of the evolution of the scalar
distance $R$ between departing photofragments and is particularly
important at longer ranges: to the best of our knowledge, no other
observable is sensitive to evolution in this large-amplitude region of
unimolecular dynamics. A key assumption of statistical theories of
unimolecular reaction dynamics is that phase space is uniformly filled
on time scales short compared to dissociation. The present results,
however, indicate that the filling of phase space takes time and is
not uniform. The present work argues that the time evolution of $R$ as
provided by TR-CEI yields a novel observation of phase space dynamics
and serves as a meaningful route to discerning the parallel
non-statistical and statistical aspects - which each participate here
- of the unimolecular dissociation of NO$_2$.\\

\section*{Data Availability}
The codes and data for the present study are available from
\url{https://github.com/MMunibas/NN-coulomb} upon publication.

\section*{Acknowledgment}
The authors gratefully acknowledge financial support from the Swiss
National Science Foundation through grants $200020\_219779$ (MM),
$200021\_215088$ (MM), and the University of Basel (MM). A. Stolow and
A. Staudte thank the NSERC Discovery Grant program, NRC QSP Challenge
Program Project QSP-075-01, and the Joint Centre for Extreme Photonics
for financial support. A. Stolow thanks the Canada Research Chairs
program for support.\\

\clearpage

\renewcommand{\thepage}{S\arabic{page}}
\renewcommand{\thetable}{S\arabic{table}}
\renewcommand{\thefigure}{S\arabic{figure}}
\renewcommand{\theequation}{S\arabic{equation}}
\renewcommand{\thesection}{S\arabic{section}} 
\setcounter{figure}{0}  
\setcounter{section}{0}  
\setcounter{table}{0}
\setcounter{page}{1}

\section*{Supporting Material: Observing Phase Space Evolution During Unimolecular Decay}

\section{Details about PESs}
In the past a wide range of potential energy surfaces (PESs) for the
reactive O+NO system has been generated and used for various
purposes. For example, the vibrational state-dependent cross sections
have been calculated from dynamics
studies.\cite{Caridade2004,Duff1994,Ramachandran2000} One
PES\cite{Duff1994} for the $^2$A$'$ state used a fit to electronic
structure calculations at the complete active space SCF (CASSCF) level
followed by multireference contracted configuration interaction and a
modified Duijneveldt (11s6p) basis set.\cite{walch:1987} Another PES
was based on 1250 (for the $^2$A$'$) and 910 ($^4$A$'$) CASPT2
calculations and fitted to an analytical function.\cite{Says2002} Such
an approach was also used for the $^2$A$''$ state.\cite{Gonzlez2001}
Complementary, a PESs for the $^2$A$'$ state using a diatomics in
molecules (DIM) expansion with the two-body terms based on extended
Hartree-Fock calculations.\cite{varandas:2003} Then, a 2-dimensional
PES with the NO bond length fixed at its equilibrium value of 2.176
a$_0$ was determined at the icMRCI+Q level of theory and a cc-pVQZ and
represented as a cubic spline.\cite{Ivanov2007} This work also
presented a PES for the $^2$A$''$ state. More recently, a double many
body expansion fit to 1700 points at the MRCI/aug-cc-VQZ level of
theory for the $^2$A$''$ state was carried out.\cite{Mota2012} In
addition, quasi classical trajectory (QCT)
calculations\cite{Caridade2004,CastroPalacio2014,Bose1997,Says2002}
have been reported for the temperature dependent rate for the
N($^4$S)+O$_{2}$ $\rightarrow$ NO+O and its reverse reaction using
different PESs. Most recently, two machine learning-based PESs were
reported. One of them used CASPT2 reference data and was represented
as a many body permutationally invariant polynomial\cite{varga:2021} and
the other one used MRCI+Q calculations as the reference method and a
reproducing kernel Hilbert space (RKHS) for representing
them.\cite{MM.no2:2020,MM.rkhs:2017}\\

\clearpage

\section{Figures}

\begin{figure}
    \centering
    \includegraphics[width=0.7\linewidth]{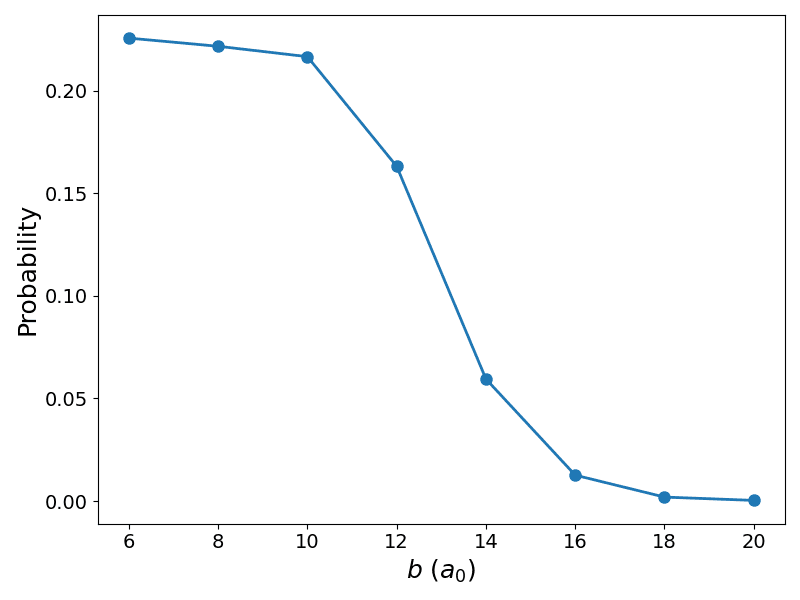}
    \caption{Opacity function for atom exchange reaction O$_{\rm A}$ +
      NO$_{\rm B}$ ($v=0, j=1, T=50$ K) $\longrightarrow$ O$_{\rm B}$
      + NO$_{\rm A}$ from QCT with initial separation $r_{\rm 0}$ =
      22 a$_0$.}
    \label{sifig:opacity}
\end{figure}

\begin{figure}
    \centering
    \includegraphics[width=1.0\linewidth]{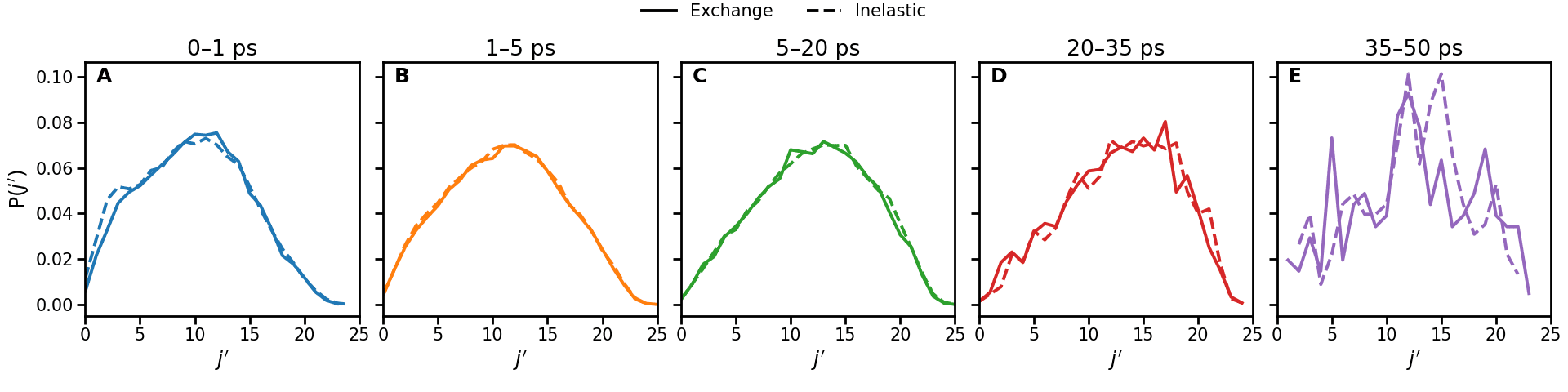}
    \caption{Final rotational distributions $P(j')$ for exchange (solid)
        and inelastic (dashed) for particular lifetime intervals.}
    \label{sifig:pofj-fin}
\end{figure}

\begin{figure}
    \centering \includegraphics[width=1.0\linewidth]{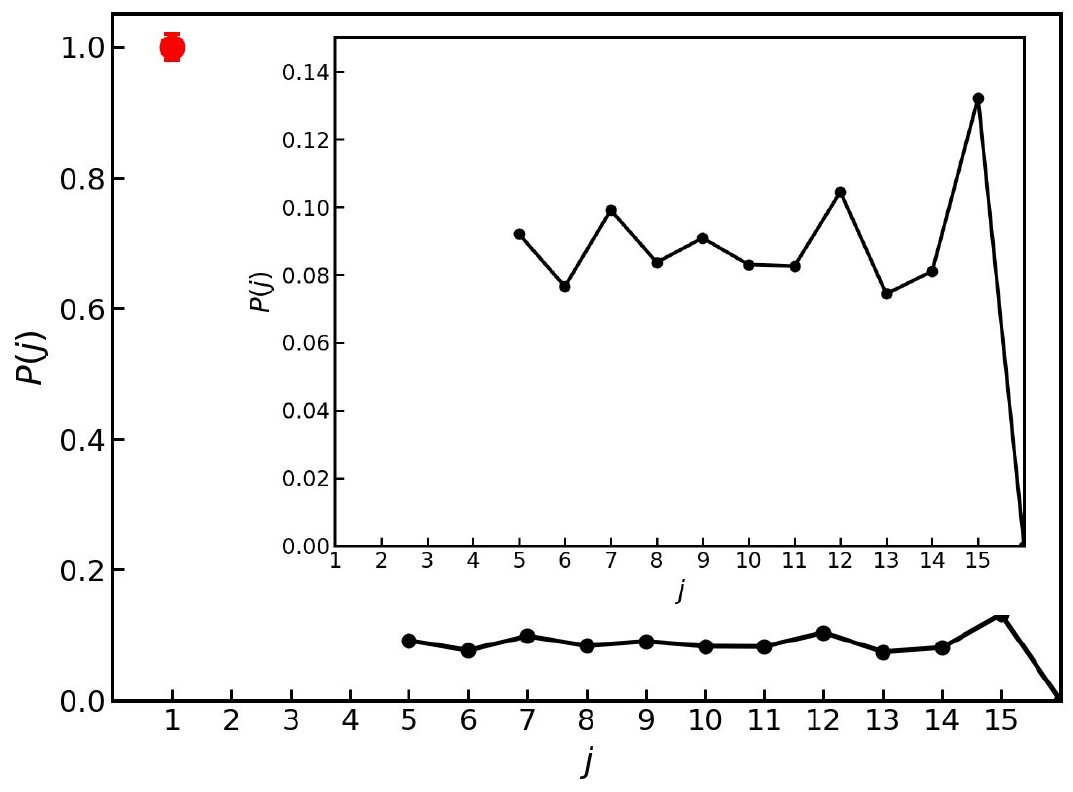}
    \caption{Comparison of the two initial rotational-state
      distributions, $P(j)$. The red circle represents the initial
      stratified sampling for impact parameter $(v = 0,\, j = 1)$,
      while the black curve corresponds to the experimentally measured
      rotational-state distribution $P(j)$ for $(v = 0,\, j =
      5\text{--}15)$, see Ref.\cite{reisler:1994}, Figure~3(a).}
    \label{sifig:pofj-ini}
\end{figure}

\begin{figure}[h!]
    \centering \includegraphics[width=1.0\linewidth]{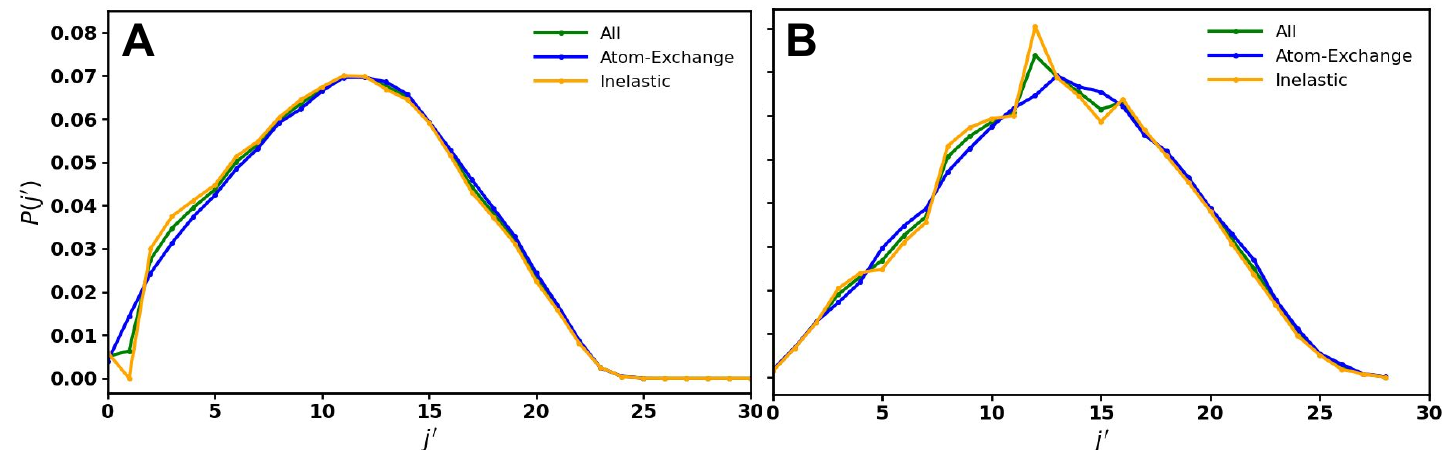}
    \caption{Final $P(j')$ probability distributions for "all"
      (green), "exchange" (blue), and "inelastic" (yellow)
      trajectories. Panel A: initial stratified sampling for NO$(v =
      0,\, j = 1)$ from 216249 trajectories analyzed in total. Panel
      B: Using the measured $P(j)$ distribution NO$(v = 0, j = 5 -
      15)$ (Figure 3a from Ref. \cite{reisler:1994}) and 122068
      analyzed trajectories.}
    \label{sifig:pofj-reisler}
\end{figure}

\begin{figure}[h!]
    \centering
    \includegraphics[width=0.8\linewidth]{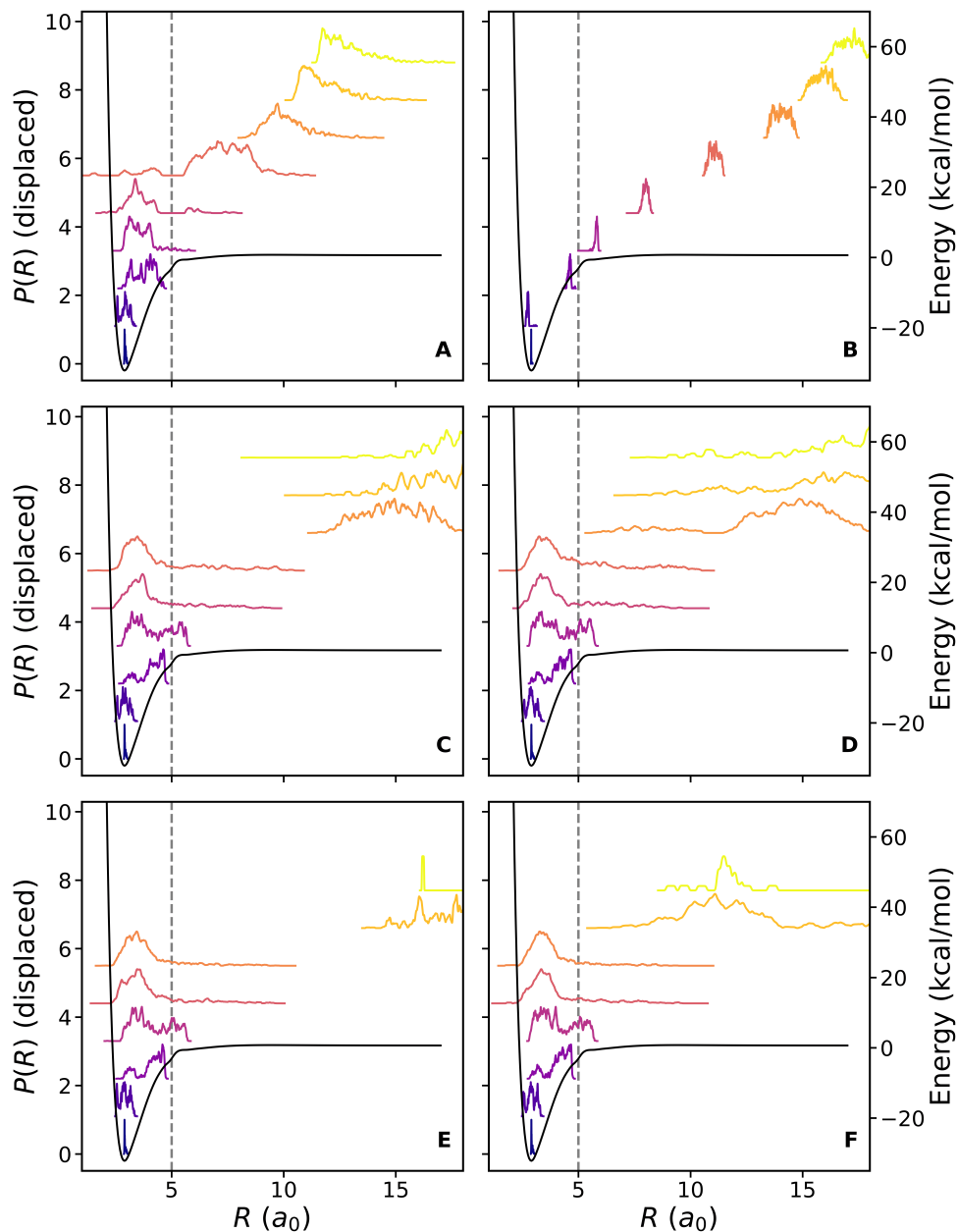}
    \caption{Analysis of $P(R)$ for exchange (left column) and
      inelastic (right column) trajectories with different NO$_2$
      lifetimes of $0.25-0.5$ ps, $2-3$ ps, and $10-15$ ps (from top
      to bottom). The curves correspond to different, $\tau$, after
      synchronization ($\tau = 1$ fs, 5 fs, 25 fs, and 50 fs, followed
      by 25\%, 50\%, 75\%, 90\%, and 100\% of the respective
      lifetimes) and are vertically offset for clarity. The vertical
      grey dashed line at $R=5$ a$_0$ shows the region of the loose TS
      inferred from $P(R,\theta)$, see Figures \ref{fig:pofrtheta1}
      and \ref{fig:pofrtheta2}. Each distribution is constructed from
      $1000$ trajectories.}
    \label{sifig:pofr-time}
\end{figure}

\begin{figure}
    \centering
    \includegraphics[width=1.0\linewidth]{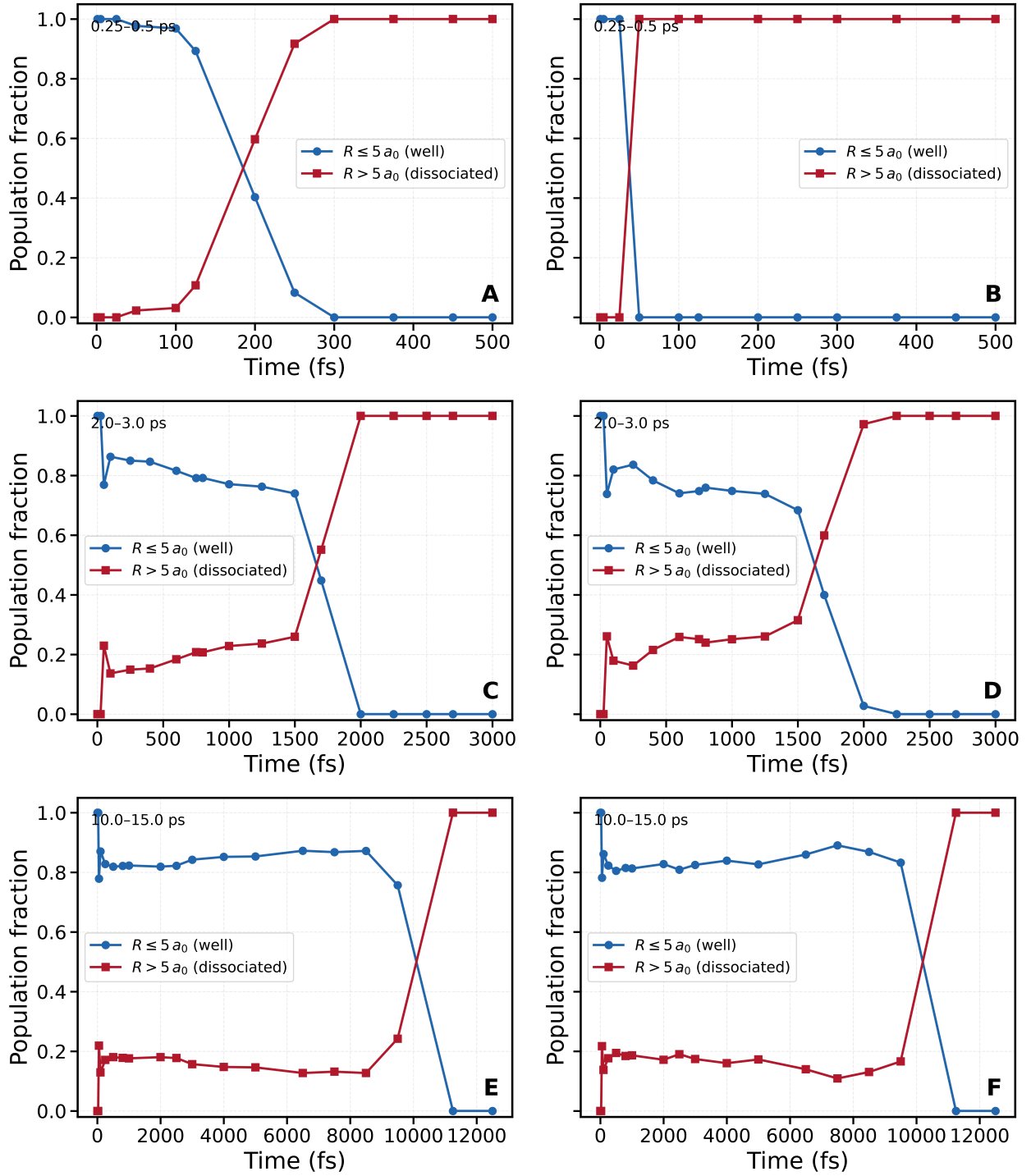}
    \caption{Integrated probabilities $\int_0^5 P(R) dR$ (blue) and
      $\int_5^\infty P(R) dR$ (red) from Figure \ref{sifig:pofr-time}
      as a function of synchronization time. Results are shown for
      three lifetime windows, $0.25$--$0.5$, $2.0$--$3.0$, and
      $10.0$--$15.0$ ps (from top to bottom). Left column for atom
      exchange and right column for inelastic scattering.}
    \label{sifig:pofr-area-1}
\end{figure}

\begin{figure}
    \centering
    \includegraphics[width=1.0\linewidth]{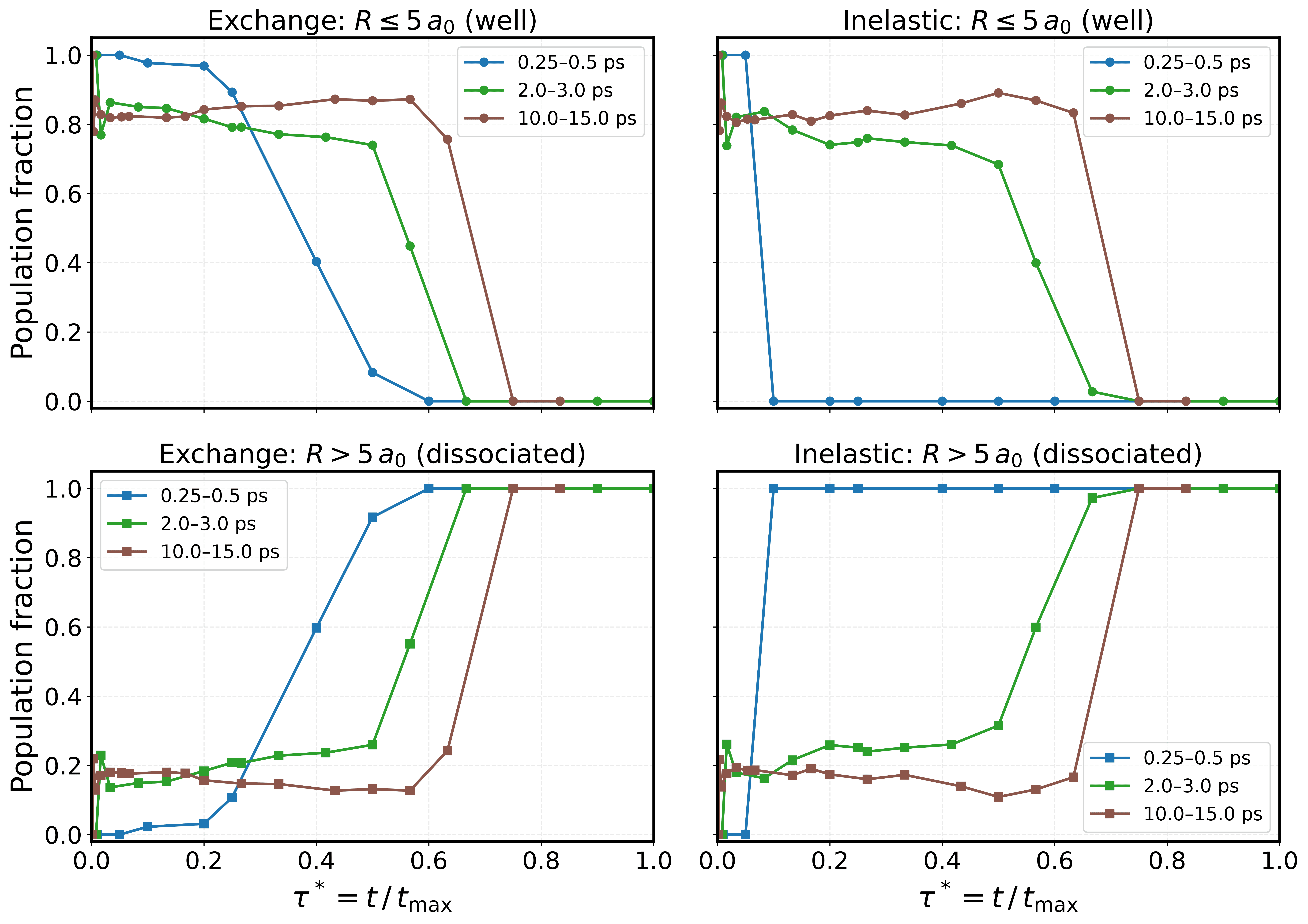}
    \caption{Integrated probabilities from Figure
      \ref{sifig:pofr-area-1} versus reduced time $\tau^* =
      t/t_{\mathrm{max}}$ for exchange (A, C, E of Figure
      \ref{sifig:pofr-area-1}) and inelastic (B, D, F of Figure
      \ref{sifig:pofr-area-1}) collisions for lifetime intervals
      0.25--0.5, 2.0--3.0, and 10.0--15.0~ps. Top: bound population
      ($R \leq 5\,a_0$); bottom: dissociated population ($R > 5\,a_0$)
      and two mechanisms (exchange, left column; inelastic, right
      column).}
    \label{sifig:pofr-area}
\end{figure}

\clearpage

\noindent
The Coulomb repulsion energies for the different reaction channels are
shown in Panels~C--E, corresponding to the inelastic, elastic, and
atom-exchange channels, respectively. As in Panels~A and~B, the
trajectories can be classified into two regimes, R1 and R2. In the R1
regime, the Coulomb energy decays on a timescale of $\sim 1$ ps,
whereas in the R2 regime it remains largely constant at values greater
than 4 eV during the first 1 ps. The trajectories contributing to the
R1 regime in Panel~A originate predominantly from the atom-exchange
channel (Panel~C), whereas those contributing to the R2 regime in
Panel~A arise predominantly from the inelastic channel (Panel~E). Only
approximately $1\%$ of all trajectories are elastic, resulting in the
sparse distribution observed in Panel~D.\\

\begin{figure}
    \centering
    \includegraphics[width=0.9\linewidth]{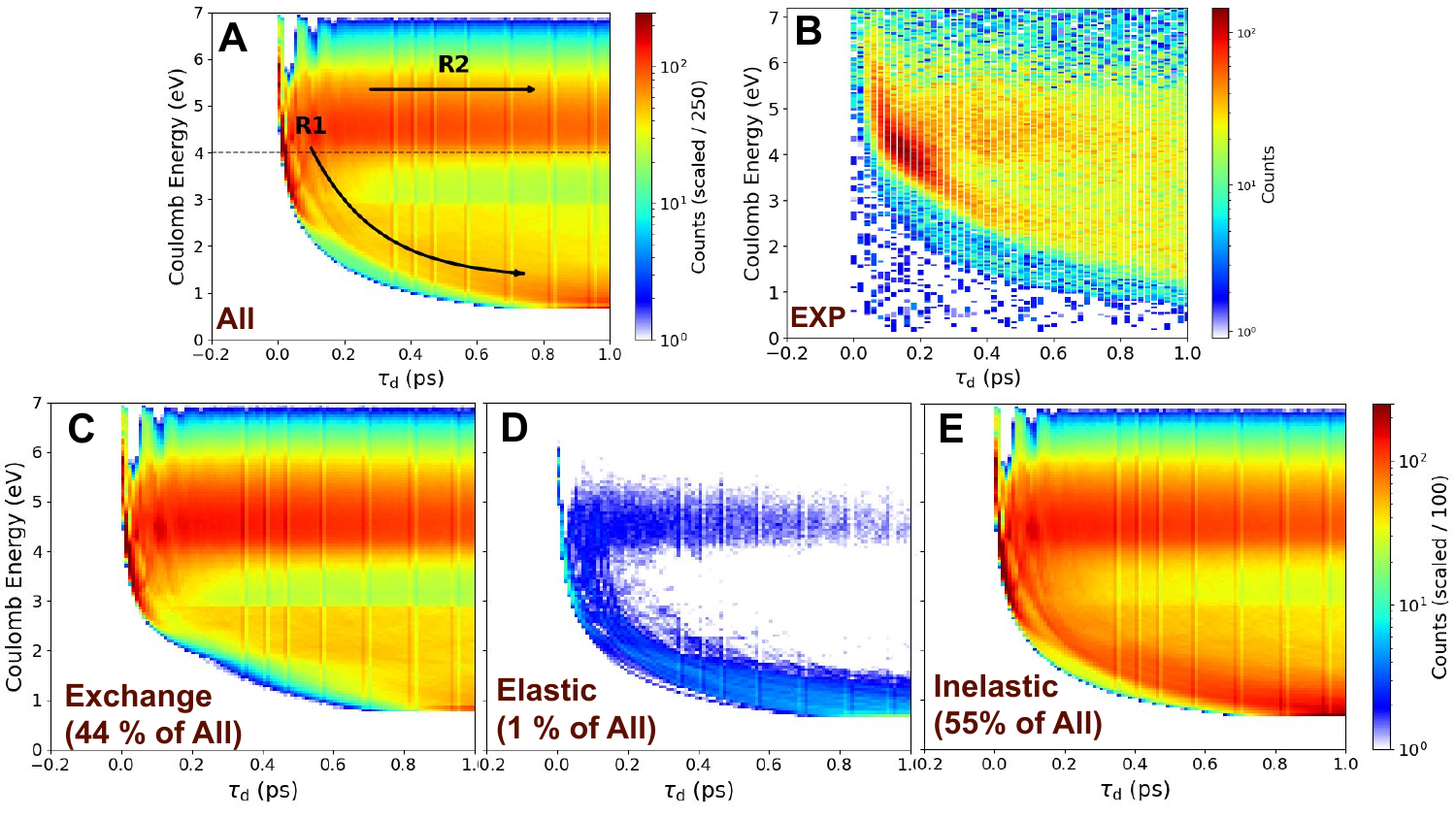}
    \caption{Coulomb repulsion energy between the \(\mathrm{NO}^{+}\)
      and \(\mathrm{O}^{+}\) cations as a function of the delay time,
      \(\tau_d\) (ps), after synchronization for 216\,249 trajectories
      that formed the \(\mathrm{NO_2}^{+}\) complex (out of 500\,000
      total trajectories). The trajectories were synchronized by
      identifying the first instance at which the distance, \(R\),
      between the centers of mass of the \(\mathrm{NO}^{+}\) and
      \(\mathrm{O}^{+}\) cations reached \(3.0\,a_0\), which defines
      \(\tau_d = 0\). The simulations employed stratified sampling of
      the impact parameter, \(b\). Panels C, D, and E show the
      contributions from atom-exchange, elastic, and inelastic
      trajectories, respectively. In Panels A, C, D, and E, the
      Coulomb repulsion energies were scaled to 60\% of their original
      values (i.e., reduced by 40\%) to place them on a scale
      comparable to the measurements.}
    \label{sifig:ecoul}
\end{figure}

\begin{figure}
    \centering
    \includegraphics[width=0.8\linewidth]{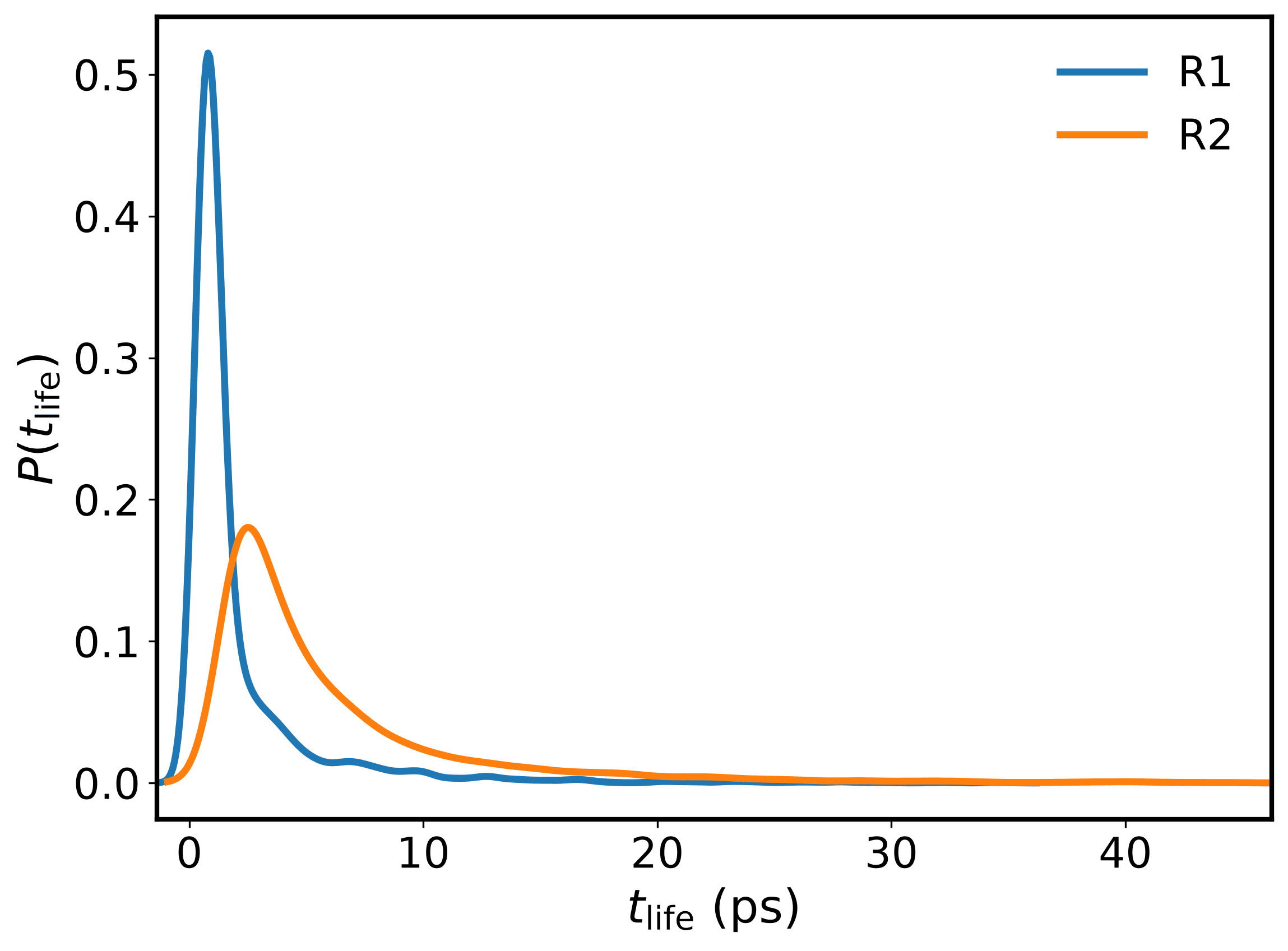}
    \caption{Distribution of lifetimes for the two regimes, R1 and R2,
      in Figure \ref{fig:ecoul}. The distributions were generated from
      3000 trajectories contributing to regimes R1 and R2, see Figure
      \ref{fig:ecoul_R1-R2}.}
    \label{sifig:life-coul}
\end{figure}

\begin{figure}
    \centering
    \includegraphics[width=0.55\linewidth]{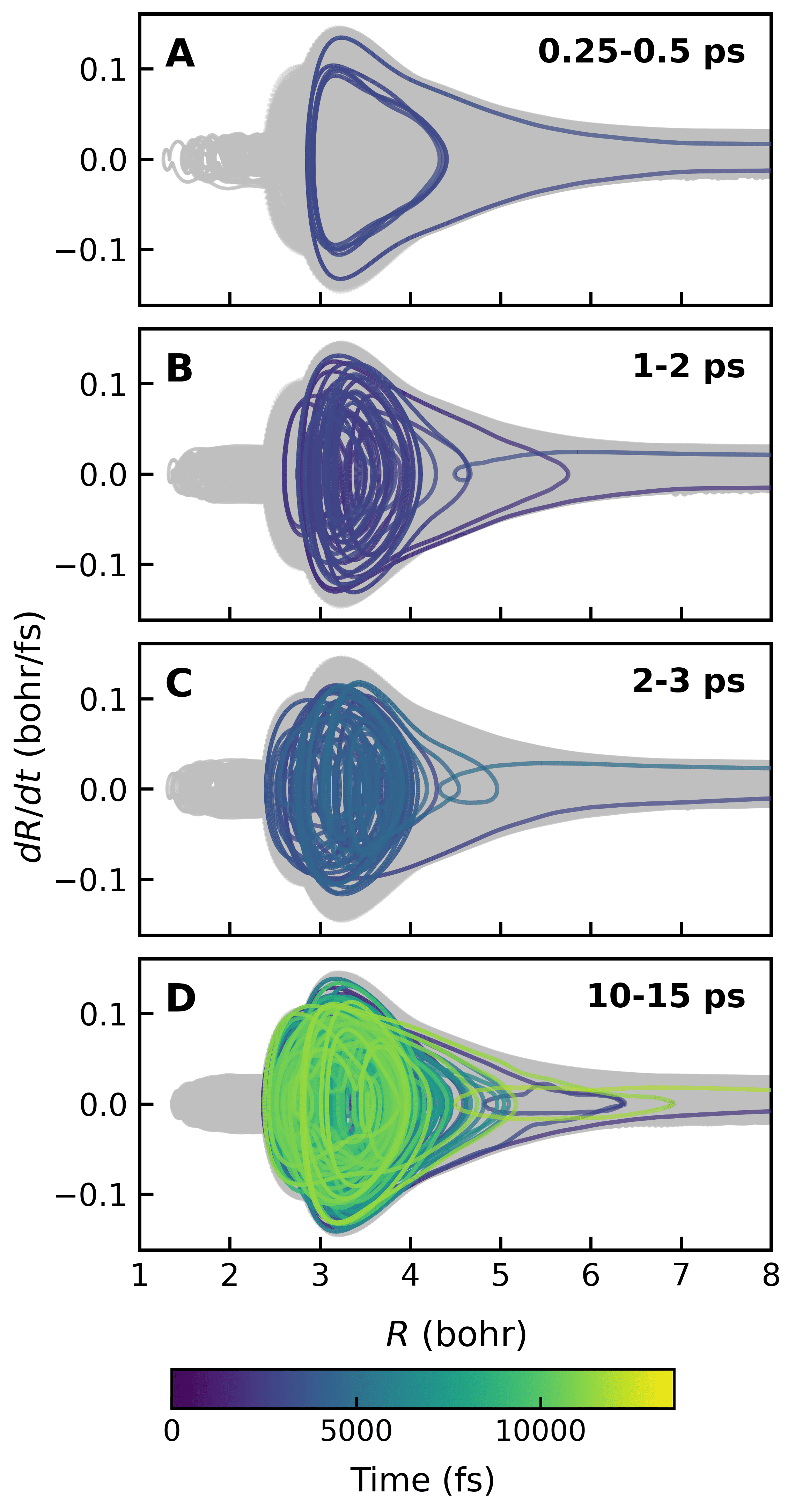}
  \caption{Phase-space representation of the radial Jacobi coordinate,
    $R$, and its conjugate velocity, $dR/dt$, for representative
    trajectories selected from different lifetime intervals: (A)
    $0.25$-$0.5$ ps, (B) $1$-$2$ ps, (C) $2$-$3$ ps, and (D) $10$-$15$
    ps. The gray areas show the corresponding ensemble of
    trajectories for each lifetime interval, whereas the highlighted
    trajectories are colored according to the elapsed time, as
    indicated by the color bar.}
    \label{fig:si_phase_R}
\end{figure}

\begin{figure}
    \centering
    \includegraphics[width=0.55\linewidth]{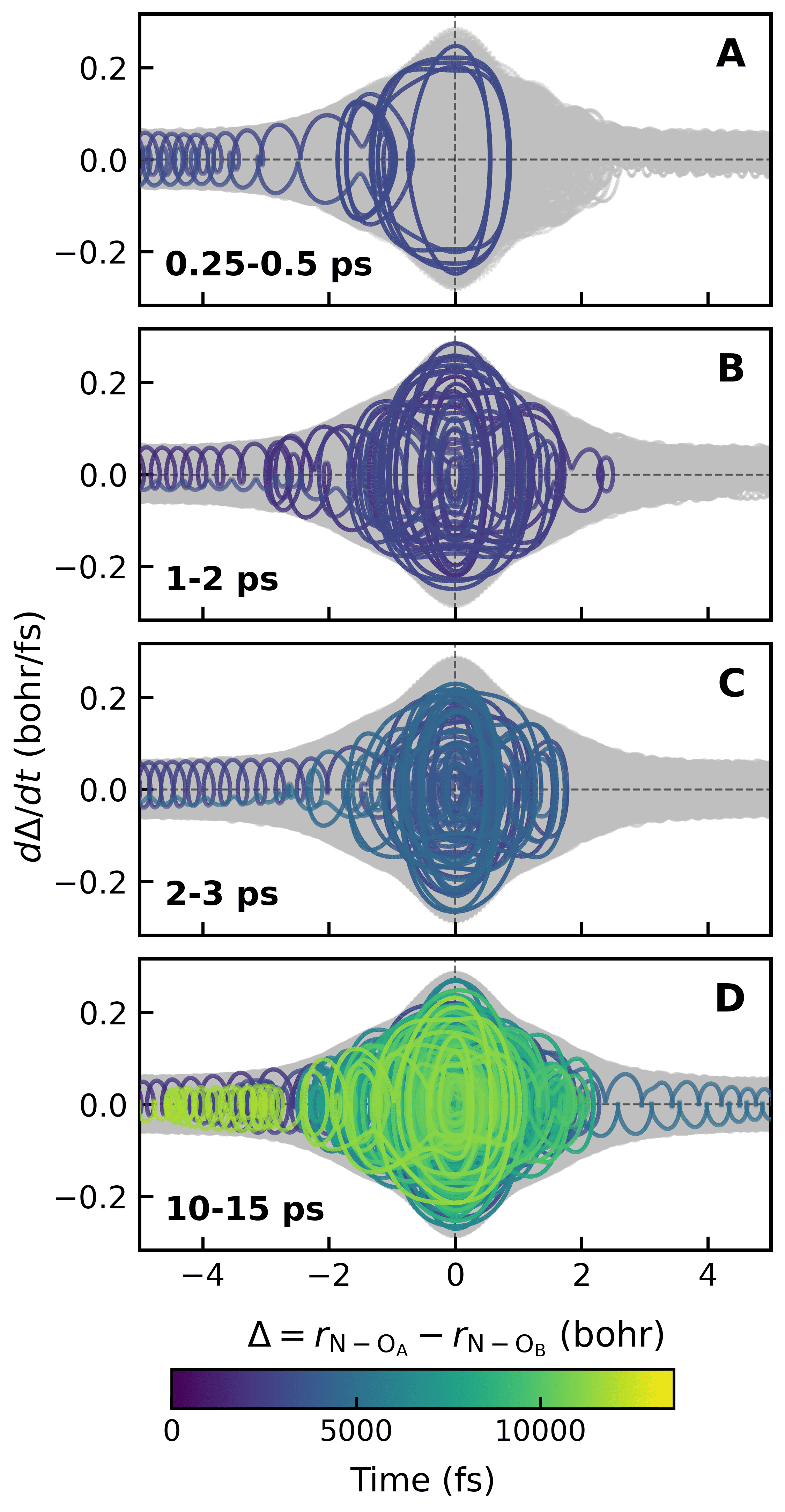}
\caption{Phase-space representation of the coordinate
  $\Delta=r_{\mathrm{N-O_A}}-r_{\mathrm{N-O_B}}$ and its time
  derivative, $d\Delta/dt$, for representative trajectories selected
  from different lifetime intervals: (A) $0.25$-$0.5$ ps, (B) $1$-$2$
  ps, (C) $2$-$3$ ps, and (D) $10$-$15$ ps. The gray areas show the
  ensemble of trajectories associated with each interval, whereas the
  highlighted trajectories are colored according to the elapsed time,
  as indicated by the color bar. The horizontal dashed line marks
  $d\Delta/dt=0$, while the vertical dashed line indicates $\Delta=0$,
  corresponding to configurations for which the two N--O distances are
  equal.}
    \label{fig:si_phase_delta}
\end{figure}

\begin{figure}
    \centering
    \includegraphics[width=0.65\linewidth]{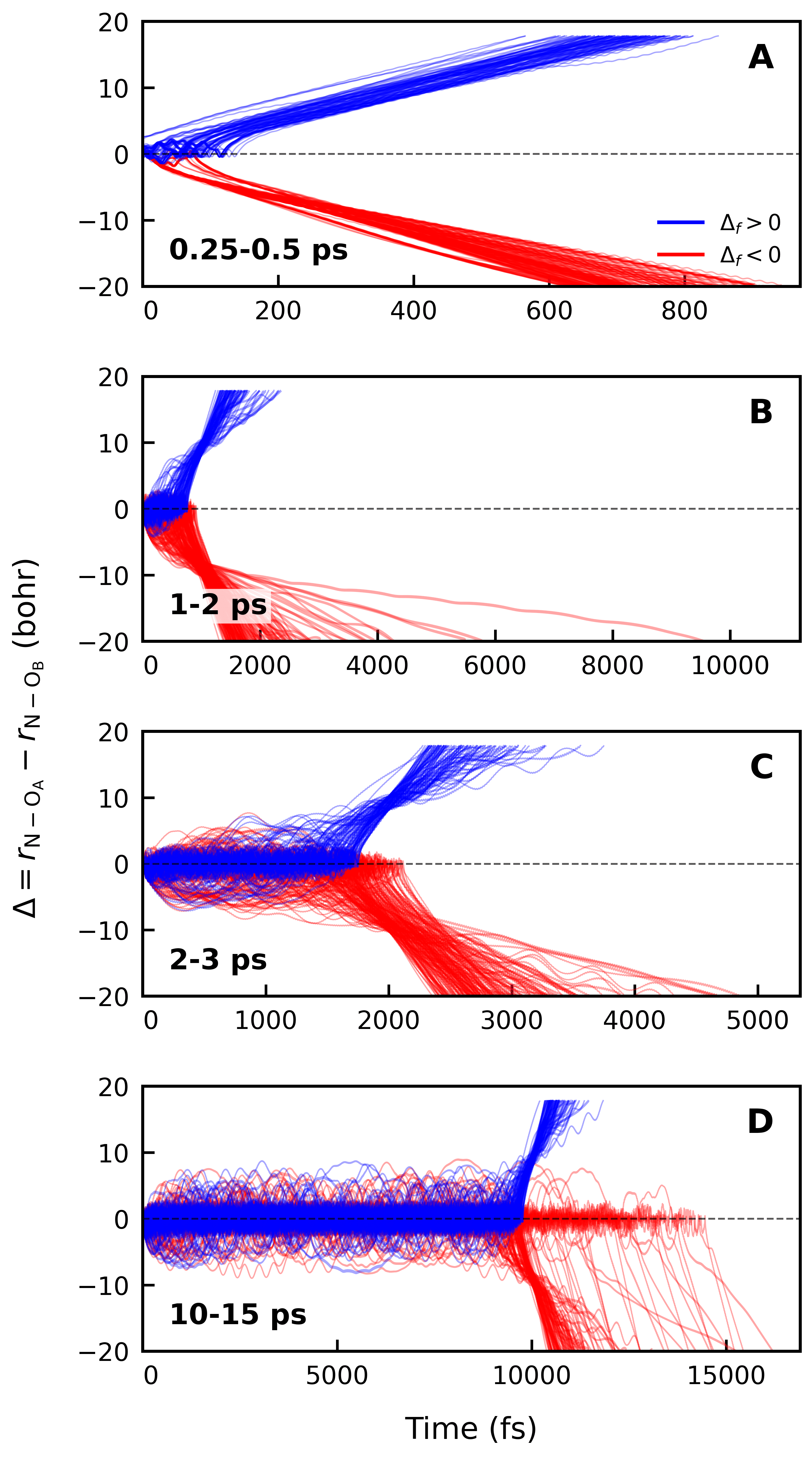}
\caption{Time evolution of the coordinate
  $\Delta=r_{\mathrm{N-O_A}}-r_{\mathrm{N-O_B}}$ for trajectories
  selected from four lifetime intervals: (A) $0.25$-$0.5$ ps, (B)
  $1$-$2$ ps, (C) $2$-$3$ ps, and (D) $10$-$15$ ps. Trajectories
  ending with $\Delta_f>0$ are shown in blue, whereas those ending
  with $\Delta_f<0$ are shown in red. The horizontal dashed line marks
  $\Delta=0$, corresponding to configurations for which the two N-O
  distances are equal.}
    \label{fig:si_delta_time}
\end{figure}

\begin{figure}[h!]
	\includegraphics[width=\textwidth]{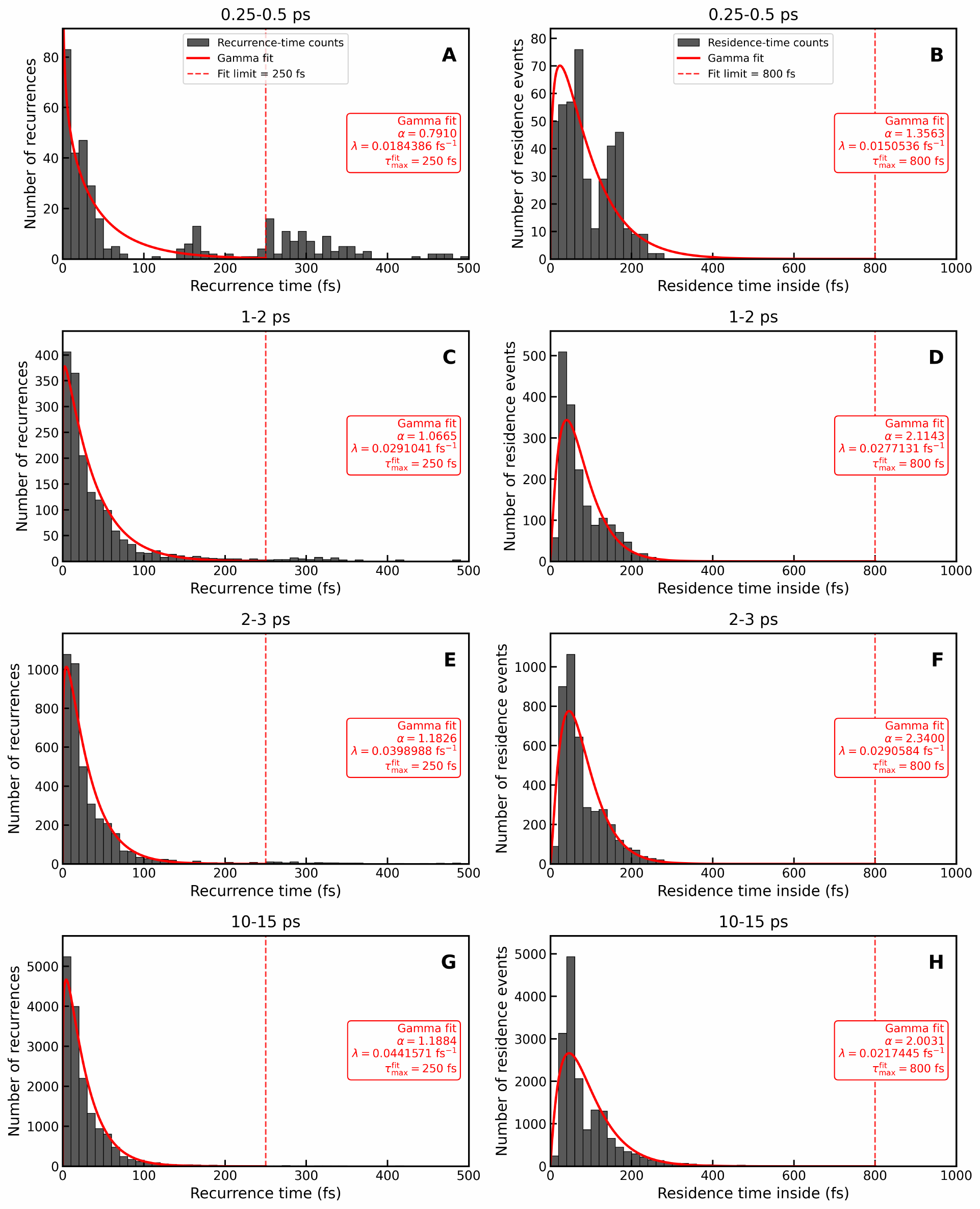}
\caption{Comparison of un-normalized recurrence-time (left column) and
  residence-time (right column) distributions for four representative
  lifetime intervals: 0.25-0.5 ps (A,B; 750 and 796 crossings), 1-2 ps
  (C,D; 2620 and 2787 crossings), 2-3 ps (E,F; 4379 and 4664
  crossings) and 10-15 ps (G,H; 16576 and 17393 crossings). For
  definitions of $t_{\rm rec}$ and $t_{\rm res}$ see Figure
  \ref{fig:sketch}. Gray histograms represent the measured event
  counts, while the red curves correspond to Gamma-distribution
  fits. The dashed vertical lines indicate the maximum recurrence or
  residence time included in each fit.}
\label{sifig:distributions}
\end{figure}

\begin{figure}
	\includegraphics[width=0.9\textwidth]{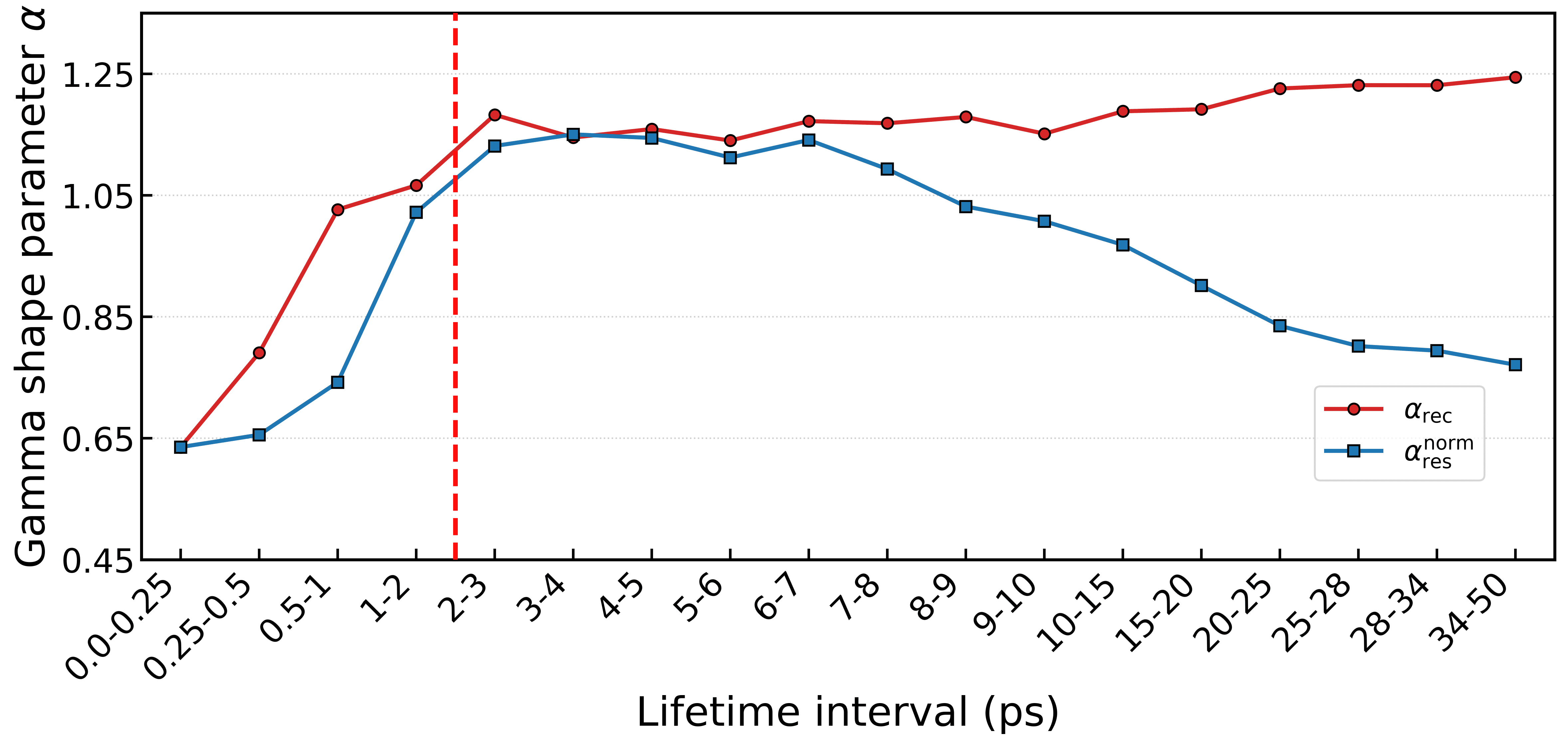}
\caption{The Gamma shape parameters $\alpha$ from Figure
  \ref{sifig:distributions} as a function of the lifetime
  intervals. The red and blue curves, report the Gamma shape
  parameters obtained from recurrence-time and residence-time
  distributions, respectively. The residence-time values are shown
  after normalization such that
  $\alpha_{\mathrm{res}}^{\mathrm{norm}}$ coincides with
  $\alpha_{\mathrm{rec}}$ in the first interval.}
\label{fig:combi_r_and_r}
\end{figure}

\begin{figure}[h!]
	\includegraphics[width=0.9\textwidth]{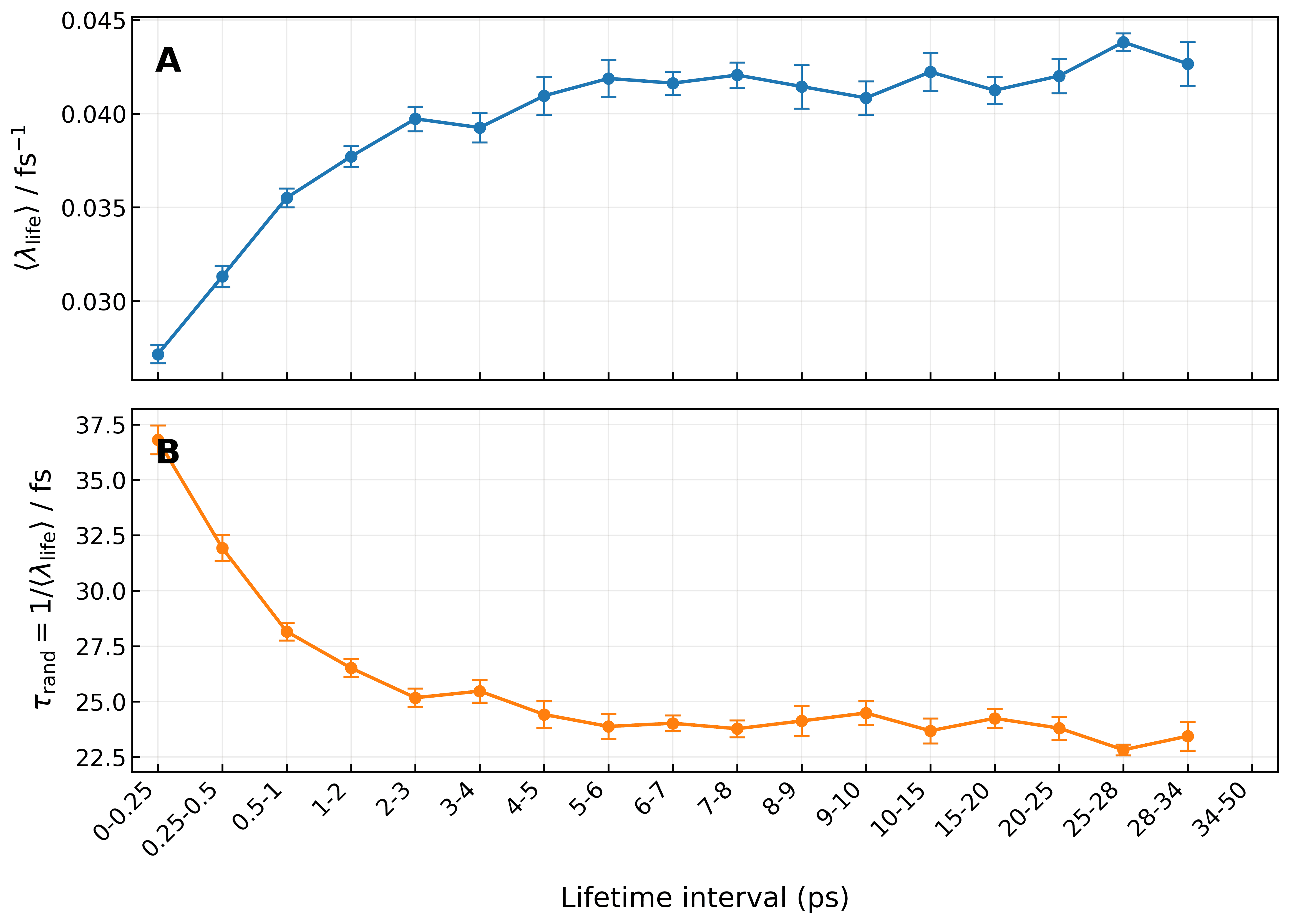}
\caption{Lifetime dependence of the Lyapunov coefficient and
  randomization time evaluated only over the collision interval.
  Panel A shows the mean absolute collision Lyapunov coefficient,
  $\langle \lambda_{\mathrm{life}} \rangle$, as a function of the
  lifetime interval, while panel B shows the corresponding
  randomization time, $\tau_{\mathrm{rand}} = 1/\langle
  \lambda_{\mathrm{life}} \rangle$.  Error bars indicate the
  statistical uncertainty associated with each lifetime interval.}
\label{sifig:lyapunov_coef_coll}
\end{figure}

\begin{figure}
    \centering
    \includegraphics[width=1.0\linewidth]{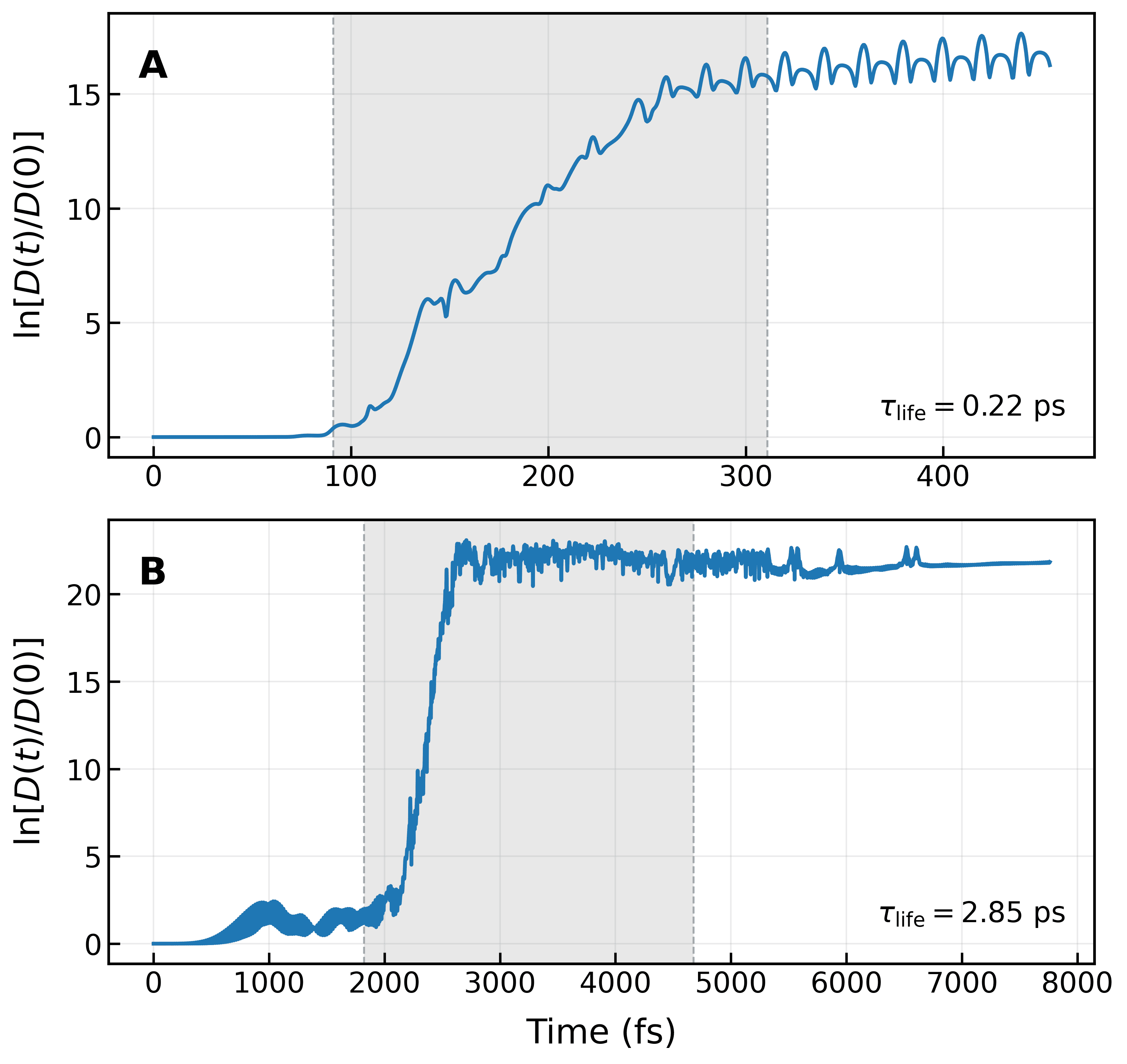}
    \caption{Time evolution of the phase-space separation between
      reference and perturbed trajectories for two representative
      collision events. The plotted quantity is $\ln[D(t)/D(0)]$,
      where $D(t)$ is the Euclidean distance between the two
      trajectories in phase space.  Panels A and B correspond to
      trajectories with lifetimes of $\tau_{\rm life}=0.22$ ps and
      $\tau_{\rm life}=2.85$ ps, respectively.  The gray shaded region
      marks the life window as defined by the lifetime criterion used
      in the trajectory calculations.  }
    \label{fig:si_distance}
\end{figure}

\clearpage

\bibliography{ref}

\end{document}